\documentclass[twocolumn]{aastex631}
\usepackage{natbib}
\usepackage{amsmath}
\usepackage{ulem}
\usepackage{booktabs}
\usepackage{graphicx}
\usepackage{graphics}
\usepackage{xcolor}

\newcommand{\github}[1]{%
   \href{#1}{\faGithubSquare}%
}

\newcommand{\je}[1]{\textcolor{black}{ #1}}
\newcommand{\jef}[1]{\textcolor{black}{ #1}}

\shorttitle{The LSSTCam: Pixel Response Characterization}
\begin{document}
\title{Photometry, Centroid and Point-Spread Function Measurements in the LSST Camera Focal Plane Using Artificial Stars}

\author[0000-0003-4373-2386]{Johnny H. Esteves}
\affiliation{Department of Physics, University of Michigan, Ann Arbor, Michigan, United States}

\author[0000-0001-6161-8988]{Yousuke Utsumi}
\affiliation{Kavli Institute for Particle Astrophysics and Cosmology, Stanford University, Stanford, California, United States}
\affiliation{SLAC National Accelerator Laboratory, Menlo Park, California, United States}

\author[0000-0002-2343-0949]{Adam Snyder}
\affiliation{Department of Physics and Astronomy, University of California/Davis, Davis, California,
United States}

\author[0000-0002-7187-9628]{Theo Schutt}
\affiliation{Kavli Institute for Particle Astrophysics and Cosmology, Stanford University, Stanford, California, United States}
\affiliation{SLAC National Accelerator Laboratory, Menlo Park, California, United States}

\author[0000-0001-6966-5316]{Alex Broughton}
\affiliation{Department of Physics \& Astronomy, University of California/Irvine, Irvine, California, United States}

\author{Bahrudin Trbalic}
\affiliation{Kavli Institute for Particle Astrophysics and Cosmology, Stanford University, Stanford, California, United States}

\author[0000-0003-3519-4004]{Sidney Mau}
\affiliation{Kavli Institute for Particle Astrophysics and Cosmology, Stanford University, Stanford, California, United States}

\author{Andrew Rasmussen}
\affiliation{Kavli Institute for Particle Astrophysics and Cosmology, Stanford University, Stanford, California, United States}
\affiliation{SLAC National Accelerator Laboratory, Menlo Park, California, United States}

\author[0000-0002-2598-0514]{Andrés A. Plazas Malagón}
\affiliation{Kavli Institute for Particle Astrophysics and Cosmology, Stanford University, Stanford, California, United States}
\affiliation{SLAC National Accelerator Laboratory, Menlo Park, California, United States}

\author[0000-0001-5326-3486]{Andrew Bradshaw}
\affiliation{Kavli Institute for Particle Astrophysics and Cosmology, Stanford University, Stanford, California, United States}
\affiliation{SLAC National Accelerator Laboratory, Menlo Park, California, United States}

\author[0000-0001-5326-3486]{Stuart Marshall L.}
\affiliation{Kavli Institute for Particle Astrophysics and Cosmology, Stanford University, Stanford, California, United States}
\affiliation{SLAC National Accelerator Laboratory, Menlo Park, California, United States}

\author[0000-0002-5296-4720]{Seth Digel}
\affiliation{Kavli Institute for Particle Astrophysics and Cosmology, Stanford University, Stanford, California, United States}
\affiliation{SLAC National Accelerator Laboratory, Menlo Park, California, United States}

\author[0000-0001-5738-8956]{James Chiang}
\affiliation{Kavli Institute for Particle Astrophysics and Cosmology, Stanford University, Stanford, California, United States}
\affiliation{SLAC National Accelerator Laboratory, Menlo Park, California, United States}

\author[0000-0001-9376-3135]{Eli Rykoff}
\affiliation{SLAC National Accelerator Laboratory, Menlo Park, California, United States}

\author[0000-0003-1989-4879]{Chris Waters}
\affiliation{Department of Astrophysical Sciences, Princeton University, Princeton, New Jersey, United States}

\author[0000-0001-6082-8529]{Marcelle Soares-Santos}
\affiliation{Department of Physics, University of Michigan, Ann Arbor, Michigan, United States}

\author[0000-0001-5326-3486]{Aaron Roodman}
\affiliation{Kavli Institute for Particle Astrophysics and Cosmology, Stanford University, Stanford, California, United States}
\affiliation{SLAC National Accelerator Laboratory, Menlo Park, California, United States}

\begin{abstract}
The Vera C. Rubin Observatory's LSST Camera (LSSTCam) pixel response has been characterized using laboratory measurements with a grid of artificial stars. 
We quantify the contributions to photometry, centroid, point-spread function size, and shape measurement errors due to small anomalies in the LSSTCam CCDs. The main sources of those anomalies are quantum efficiency variations and pixel area variations induced by the amplifier segmentation boundaries and ``tree-rings" — circular variations in silicon doping concentration. 
This laboratory study using artificial stars projected on the sensors shows overall 
small effects. % , residuals below $0.1\%$, \jef{except near} the CCD mid-line. 
The residual effects on point-spread function (PSF) size and shape are below $0.1\%$, meeting the ten-year LSST survey science requirements. However, the CCD mid-line presents distortions that can have a moderate impact on PSF measurements. 
This feature can be avoided by masking the affected regions. Effects of tree-rings are observed on centroids and PSFs of the artificial stars and the nature of the effect is confirmed by a study of the flat-field response. Nevertheless, further studies of the full-focal plane with stellar data should more completely probe variations and might reveal new features, e.g. wavelength-dependent effects. The results of this study can be used as a guide for the on-sky operation of LSSTCam.

% We studied the effects  using artificial stars projected on  the sensors and find that  the resulting  measurement uncertainties pass the ten-year LSST survey science requirements. In addition, we verify that the tree-ring effects can be corrected using flat-field images if needed, because the astronomic shifts and shape measurement  errors they induce correlate well with the flat-field signal. Nevertheless, further sensor anomaly studies with on-sky data should probe possible temporal and wavelength-dependent effects. 
\end{abstract}

% \tableofcontents

\section{Introduction} \label{sec:intro}

The Vera C. Rubin Observatory is a next-generation optical and near-infrared observatory currently under construction in Cerro Pachón, Chile. The Rubin Observatory will conduct the Legacy Survey of Space and Time (LSST), an unprecedented galaxy survey of 18000 sq-deg of the southern sky that will revisit each area over 825 times in 10 years and in six photometric bands, \emph{ugrizy}. The four science pillars of  LSST main are to probe the nature of dark energy and dark matter, take an inventory of the solar system, explore the transient optical sky, and study the evolution and structure of the Milky Way \citep{ivezic19}. 
To achieve these goals, the Rubin Observatory will use the 3.2 gigapixel LSST Camera (LSSTCam), mounted on the 8.4-meter Simonyi Survey Telescope. The LSSTCam has a large field of view of approximately 10 sq-deg, with a focal plane of 201 4k by 4k, thick (100 $\mu m$), fully-depleted, back-illuminated charge-coupled devices \citep[CCDs;][]{holland03,holland09,holland14}. 

The focal plane of the LSSTCam is populated by 189 science CCDs, 8  CCDs for auto-guiding, and 4 split CCDs for wavefront measurements. The focal plane consists of 25 sub-assemblies, 21 Science Rafts \citep{O'Connor16} with a 3$\times$3 mosaic of science CCDs and 4 Corner Rafts \citep{Arndt10} each with two guiders and one split wavefront sensor. For LSST custom CCDs, \je{an array of $(10 ~{\rm \mu m})^2$ pixels with a thickness of} 100 $\mu m$ back-illuminated deep-depletion devices were developed. These sensors feature 16 amplifier segments, each 2k by 0.5k, arranged in two rows of eight segments and separated by a mid-line break, to enable low-noise, two-second readout via parallel readout of the segments. The CCD sensors were fabricated by two vendors: Imaging Technology Laboratory (ITL) and Teledyne e2v (e2v). Both vendors meet the LSST CCD requirements \citep{2009JInst...4.3002R,2014SPIE.9154E..18D,2016SPIE.9915E..0VK} but there are subtle differences between those sensors. %, as well as from sensor to sensor from each vendor. 

Thick, high-resistivity CCDs have been used by other wide-field imagers such as the Dark Energy Camera (DECam, \citet{flaugher15}) and the Hyper Suprime-Cam (HSC, \citet{miyazaki18}), in part due to their high quantum efficiency (QE) at longer wavelengths (near infrared). However, these types of detectors have been found to imprint subtle but significant undesirable characteristics that impact centroid, photometric, flux, and shape measurements \citep{stubbs14,Astier2015,mandelbaum2015}. Study and characterization of any source of systematic errors will be crucial to achieving the required accuracy to achieve the scientific goals of a survey such as the LSST.

An initial study characterizing prototype LSST sensors was performed to ensure that the CCDs met basic LSST performance requirements for, e.g., read noise, quantum efficiency, charge transfer efficiency, diffusion and full well \citep{2014SPIE.9154E..18D}. \cite{2018SPIE10709E..2BS} studied optimization of operational voltages for sensors from ITL. \cite{2021JATIS...7d8002S,Snyder_2021} performed measurements of effects of sensor anomalies such as deferred charge distortions, centroid shifts, and the brighter-fatter effect \citep{Gruen2015}. \citet{Park_2017,Park_2020} studied the imprints of circular patterns due to silicon dopant concentration variation (tree-rings effect) on flat-field images for the full set of LSST CCDs. 

\jef{In addition, \citet{Juramy_2020} studied an anomaly called ``tearing", a visually striking distortion created during the e2v readout.  The distortion at the mid-line break and the amplifier boundaries is different dynamically with respect to light.} The bias and clock voltages as well as the CCD controller sequence were optimized in the course of individual Raft and focal plane testing. Comprehensive results of the overall testing campaign will be described in a future publication. 
% \xout{In addition, we anticipate that full characterization of each individual sensor will be necessary to achieve the desired level of precision from the LSSTCam}\citet{Bond18,Roodman18,2021JATIS...7d8002S}

% \je{This paper is a product of a image campaign to evaluate the uniformity of the photometry, centroid and measured shapes of artificial stars. It verifies the capability of an individual LSSTCam sensor in conducing the 10 year survey as outlined by the LSST requirements. Also, it presents a summary of the expected astrometry, photometry and PSF measurements distortions. }

\je{This paper presents a initial assessment of sensor effects \jef{related to moments of the brightness distributions (up to second order) of star-like spots}, for selected sensors. We quantify the observed anomalies in terms of the desired \jef{limits on systematic uncertainties for the LSST}, paying special attention to effects occurring at certain locations on the CCDs. \jef{Effects from tree-rings are} observed \jef{in the measured} centroid and point-spread function (PSF), and in flat-field response, \jef{with a consistent interpretation of the origin. Our results} will be the foundation for further investigation of the sensor systematics over the full focal plane  using on-sky data, and development of future corrections\jef{, if needed.  However,} these topics are beyond the scope \jef{of the present work.}}

\section{Focal Plane: Spot Grid Test}\label{sec:section2}
This section first describes the laboratory test setup and the data acquisitions, then the imaging data collection and the post-processing methodology, which includes the source detection, the grid fitting algorithm, and the calibration of the spot-measured quantities: flux, centroid, PSF shape, and size.  

\subsection{Camera Bench For Optical Testing: Spot Projector}\label{sec:test-setup}
The focal plane in the Camera cryostat was mounted on the top of the Bench for Optical Testing (BOT), facing downward. The BOT was designed to achieve a dark environment for electro-optical testing of the focal plane. The background light is reduced to below 0.01 electrons per second per pixel, an ideal environment to perform optical tests without light contamination. For a complete description of the BOT design, assembly, and requirements, see \cite{Newbry_2018, Snyder_2021}.

Underneath the BOT, we mounted the spot projector, which creates the artificial star grid \jef{by projecting the image of a spot mask onto the focal plane.}
% The spot projector comprises a light source, a shutter, an integrating sphere, a photographic mask on a controllable filter wheel, and a commercial lens. In particular, the shape and size of the artificial stars are set by the photographic spot mask characteristics. 
The spot projector experimental apparatus has a 450 nm light-emitting diode (LED) light source fed by an optical fiber into the integrating sphere gated by a single-blade beam shutter (``Thorlabs 1”). The spot pattern is set by a photographic mask (HTA Photomask photo-lithographic) on the filter wheel. The commercial lens (Nikon 105mm f/2.8 Al-s Micro-Nikkor) is used to re-image the integrating sphere's 1” exit port. The F-stop ring was set to be closed as much as possible. The entire image of the mask is about the size of an LSST CCD.

\begin{figure*}[!ht]
    \centering
    \includegraphics[width=0.9\textwidth]{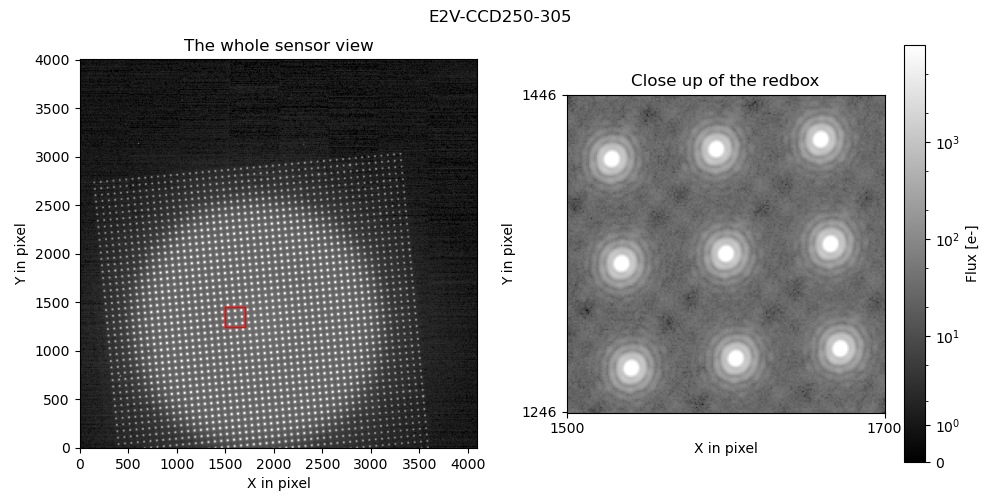}
    \caption{Example images of (left) the 49x49 spot grid and (right) a close up of the highlighted region on the left, respectively. The star-like spots are approximate point sources, with $\text{FWHM}$ 5.2 pixels. The grid allow probing much of the sensor area with each acquisition. Note that the sources located in the corners are masked due to their diminished flux, a result of the vignetting effect produced by the commercial lens in our setup.
    }\label{fig:spotgrid_illustration}
\end{figure*}

% The magnification of the system is close to 3, the pixel scale on (75 pixels between spots on the focal plane, 0.25mm spacing at the mask) the final beam is about F/100 in the image plane.
 The spot projector is placed on a remotely controlled XY stage. This XY stage allows the translation of the projector to point at any location of the focal plane. \je{The spots projected onto the focal plane form a uniform, rectangular grid with 750~$\mu m$ (75~pixel) spacing between nearest neighbors. Overall, the grid \jef{has} 49$\times$49 spots, as shown in} \autoref{fig:spotgrid_illustration}.

\subsection{Data Acquisition}
We collected images for this analysis in two series, Run 3 and Run 5. We began with a randomly selected sensor from each vendor: R22-S11 (e2v) and R02-S02 (ITL). \je{It was a pilot study during the third} round of electro-optical testing campaigns at SLAC (Run 3; 2019/10/4--2019/11/5). Then we extended the scope to four additional sensors in Run 5 (2021/11/4 and 2022/1/6) with some improvements in the acquisition procedure: two ITL sensors (R03-S12, R10-S11) and two e2v sensors (R24-S11, R32-S01). The nominal flux of the spots was set at $\sim$50000\,e-/pixel, which is bright relative to the readout noise, $\sim$10\,e-, and well below sensor full well, $\sim$100000\,e-.
 The sensor voltages readout sequence parameters are different between those two data acquisition campaigns. The specific parameters are tabulated in \autoref{appendix:Operation}.

\begin{figure}[!ht]
    \centering
    \includegraphics[width=0.50\textwidth]{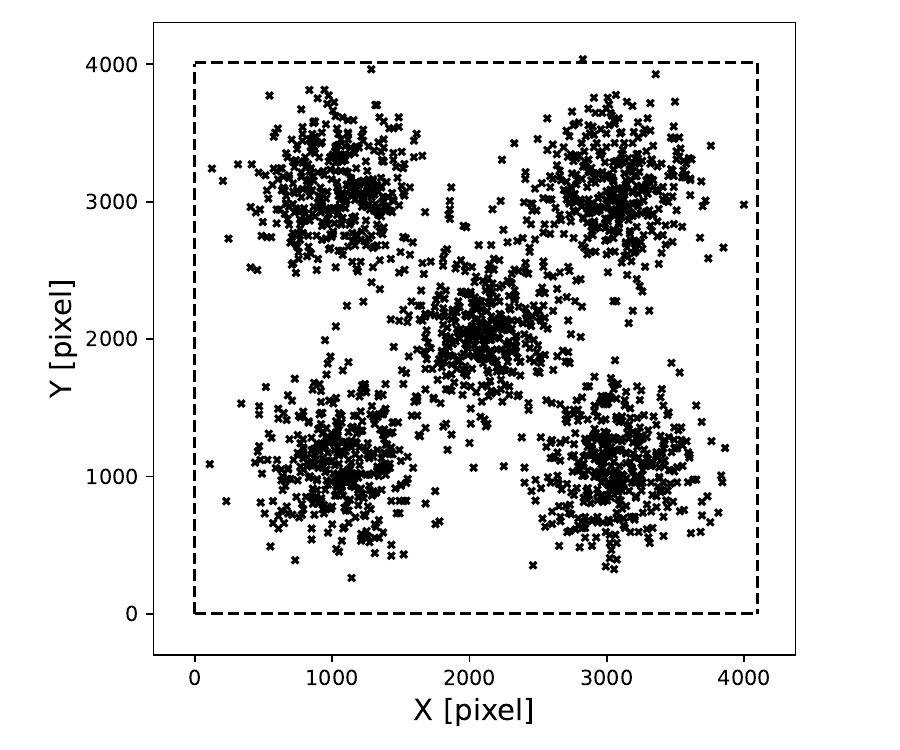}
    \caption{\je{An example of the updated dithering pattern for R24-S11. The crosses represent \jef{the locations of the centers of the spot grid pattern for the 2000 acquisition.} The CCD frame is outlined by the dashed lines.}
    }\label{fig:dithering_pattern}
\end{figure}

\jef{In Run 3,} we collected 1600 images with the spot projector position \jef{randomly} dithered around the center of each  of the studied CCDs, spanning $\pm$5 mm in Run 3. However, we found two major technical issues with the Run 3 images: (a) the projected spot grid did not cover an entire CCD; (b) the images were not sharply focused. Therefore, in Run 5, we changed the acquisition to 2000 \jef{randomly} dithered images at the center and four quadrants of the studied CCDs, spanning $\pm$ 5mm to cover the entire CCD. Figure \ref{fig:dithering_pattern} shows the dithering pattern.

For Run 5 we also replaced a manual adjustable Z-stage on the XY stage with a remotely controlled Z-stage to address the focus issue. This updated setup significantly improved the focusing process. However, the structure of the Z-stage introduced vibrations that were significant at the physical size of the projected spots. \jef{These} effects were corrected through post processing \jef{because they were identified} during data analysis after Run 5 was completed. In \autoref{sec:residuals}, we describe our corrections. 

\subsection{Artificial Stars Image Collection}\label{sec:image_collection}

An example of a single exposure taken during the imaging acquisition process is shown in \autoref{fig:spotgrid_illustration}. The left-hand panel illustrates our star-like field, a square grid with $49\times49$ optical spots projected on the CCD. Note that the grid \jef{spans} a large part of the detector in a single exposure. The star-like physical size and Gaussian shape of the spots may be discerned from the figure. The spots on the grid are equally spaced and have FWHM of $\sim5$ pixels, corresponding to $\sim 1^{\prime\prime}$ in the LSSTCam focal plane. With \jef{an average} ellipticity of $\sim 0$, the spots are quite similar to point sources, and their size was chosen to be equivalent to the PSF of the camera. 

The optical setup creates image artifacts as shown \jef{in} \autoref{fig:spotgrid_illustration}. For instance, the commercial lens vignettes the image; the spots on the edges are affected and have significantly lower flux. In addition, the spots have a characteristic shape and size that vary across the grid. The setup design can then impact our analysis if not treated correctly. Therefore, having many exposures and different spots for each CCD pixel is desirable to mitigate effects from the experimental setup. \autoref{sec:residuals} further describes our statistical treatment of the experimental setup imperfections.

\subsection{Single Image Processing And Grid Characterization}\label{sec:image_processing}
Here we describe the analysis for a single exposure, from raw image to the resulting catalog of identified sources. The spot exposures presented in \autoref{sec:image_collection} were analyzed using version 21.0.0 of the LSST Science Pipelines \cite[hereafter pipelines; \url{https://pipelines.lsst.io};][]{bosch2018,Bosch2019}. The pipelines is a set of data processing tasks actively being developed to process the LSST data. The raw images taken of the spot grid were processed using the standard pipelines instrument signature removal (ISR) task. An extension to the standard pipelines for these lab spot images was developed and used (\href{https://github.com/Snyder005/mixcoatl/}{\texttt{mixcoatl}}). This procedure includes a bias level subtraction using the row-by-row median value of the overscan region, 2D structure in the bias using a medianed overscan-subtracted bias image, masking of pixel defects, and applying gain correction as derived from \jef{measurements using an Fe$^{55}$ X-ray source. } \jef{To better observe sensor anomalies, we do not apply flat field corrections.}

After ISR processing, the identification of sources and measurement of source properties was performed using a custom source detection task \texttt{mixcoatl.characterizeSpots.CharacterizeSpotsTask}. This task detects sources by applying a maximum filter to the image and identifying peaks above a pixel value threshold of 200 electrons. This methodology was needed because the spatial variation of the projector's scattered light background was ill-suited for the background modeling and subtraction performed by the standard pipelines source detection task. \jef{The detected sources' fluxes ($f$), positions ($x,y$), and second moments of brightness ($I_{xx}$, $I_{xy}$, and $I_{yy}$) were measured using the SDSS HSM algorithm \citep{HS2003,Mandelbaum2005,bosch2018}. The second moment of brightness for an object $S(\boldsymbol{x})$ is given by: }

\begin{equation}
    I_{i j} \quad=2 \frac{\int_{\mathbf{R}^2}\left(\boldsymbol{x}-\boldsymbol{x}_0\right)_i\left(\boldsymbol{x}-\boldsymbol{x}_0\right)_j w(\boldsymbol{x}) S(\boldsymbol{x}) \mathrm{d}^2 \boldsymbol{x}}{\int_{\mathbf{R}^2} w(\boldsymbol{x}) S(\boldsymbol{x}) \mathrm{d}^2 \boldsymbol{x}} \, ,
\end{equation}
\jef{where $w(\boldsymbol{x})$ are the weights that maximizs the S/N under the assumption of an elliptical Gaussian shape. The algorithm details can be found in \citet{bosch2018}. }

% For the record, from the pipelines we used the HSM flux \citep{HS2003} ($f$: \texttt{base\_SdssShape\_instFlux}), the centroids ($\vec{\ell}$: \texttt{spotgrid\_x/y}), and the elliptical Gaussian adaptive moments ($I_{xx}$, $I_{xy}$, and $I_{yy}$: \texttt{base\_SdssShape\_xx/xy/yy}) quantities.

%
The next step was to derive the properties of the projected grid from the set of detected sources, including the overall magnification, the row/column spacing, and the rotation of the grid with respect to the pixel array. We filtered the outlier sources with the following threshold $2.0 < I_{xx} ~\text{ or}~I_{yy} < 20.0 \text{ px}^2$ in order to exclude sources that do not correspond to points on the projected grid. The remaining sources were then fit to an ideal grid model of $49\times49$ spots, with three free parameters corresponding to the x/y grid center position and $\theta$ the grid angle. We employed a least-squares minimization approach to minimize the distances between detected and grid model source positions. To initiate the grid model fitting step, we utilized a convex hull technique to provide an initial guess. The convex hull method proved to be more robust than using the commanded grid center as an initial value. After determining a best-fit model grid, each detected source was assigned an index label corresponding to its row and column number in the projected grid; this identification allowed for tracking individual sources across exposures. The per-source position residuals from the ideal grid model were then calculated and recorded in the source catalog for the measurement of optical and sensor distortions, as described in \autoref{sec:residuals}.

\subsection{Measurement Calibrations: Residuals}\label{sec:residuals}
Here we describe how we calibrate our measurements.% and address the errors introduced by the Z-stage vibration in the Run 5 setup. 

As shown in \autoref{sec:image_collection}, the spots are not ideal point sources, and their locations on the grid, shape deviations, and projector lens aberrations, for example, impact their measured quantities. In order to remove effects of the optical aberrations, we compute the residuals of a measured and ideal spot property for a large collection of exposures centered at different positions. The residual vector $\delta \vec{\ell}$ between the position of an ideal grid spot $\vec{s}$ and the position of the corresponding detected spot $\vec{d}$ is \citep{Snyder_2021}:
\begin{equation}
    \delta \vec{\ell} = \delta \vec{\ell}_{optical} + \delta \vec{\ell}_{sensor} + \vec{\epsilon} \; ,
\end{equation}
where $\vec{\epsilon}$ is a random error term, $\delta \vec{\ell}_{optical}$ represents displacements that can be caused by the optical setup, including the mask used to generate the spots \jef{that are constant and independent of the position of the projector. If we average} the residual $\delta \vec{\ell}$ of one single spot measured \jef{at many different locations on the CCD,} the sensor anomalies should average out, and only the constant displacements due to the optical setup \jef{will} remain. \jef{For a sufficiently} large set of residual measurements, the sensor anomalies should be:
\begin{equation}
    \delta \vec{\ell}_{sensor}  = \delta \vec{\ell} - \left< \delta \vec{\ell} \right> \; ,
\end{equation}
where $\left< \delta \vec{\ell} \right>$ is a constant value for one spot since it is the average residual vector over the CCD pixels. For randomly distributed errors, the expectation value of the error term is zero.  The statistical error can be reduced by increasing the sample size.

Similarly, the residuals of the PSF shape and size are computed in terms of the residuals of the second moments of brightness  \citep{Bernstein2002}:
\begin{equation}
    \delta I_{ij} = I_{ij} - \langle I_{ij} \rangle \; ,
\end{equation}

The second moments of brightness are a building block for the PSF shape and size measurements \citep{1995ApJ...449..460K, Schneider2005}: 
\begin{align}
    T   &=  I_{xx}+ I_{yy}\\
    e_1 &=  \left(I_{xx} - I_{yy}\right)/T\\
    e_2 &= 2 I_{xy}/T
    \label{eq:secondMoments-shape}
\end{align}
where $T$ is the PSF-size, and $e_1$ and $e_2$ are the $x$ and $y$ components of ellipticity. We note that others definitions of ellipticity from the second moments of brightness are possible \citep{Schneider2005}. Also, the above definition of PSF-size ($T$) relates to the PSF full width at half maximum (FWHM) as: 
\begin{equation}
    \text{FWHM} \equiv 2.355 \sigma = 2.355 \sqrt{T/2} \text{  ,}
\end{equation}
if the profile is Gaussian. 

The resulting residuals of the PSF shape and size are:
\begin{align}
    \delta T   &= \delta I_{xx}+ \delta I_{yy} \\
    \delta e_1 &= (\delta I_{xx}- \delta I_{yy})/ \langle T\rangle \\
    \delta e_2 &= 2 \delta I_{xy}/\langle T\rangle \; . 
\end{align}

We computed the fractional flux residuals in terms of the ratio:
\begin{equation}
    \frac{\delta f}{f} =  \frac{f-\langle f\rangle}{\langle f\rangle} \; ,
%    \delta m|_{\rm sensor} =  m - \overline{ m} = - 2.5 \log{\left(\frac{ f}{ f_{\rm mean}}\right)} \; 
\end{equation}
For small fluxes residuals ($\delta f/f$) can be expressed as a multiplicative factor, where $\delta m \approx -1.08573 \delta f/f$ mag. 

\jef{For the record, from the pipelines we used the HSM flux ($f$: \texttt{base\_SdssShape\_instFlux}), the centroids ($\vec{\ell}$: \texttt{spotgrid\_x/y}), and the elliptical Gaussian adaptive moments ($I_{xx}$, $I_{xy}$, and $I_{yy}$: \texttt{base\_SdssShape\_xx/xy/yy}) quantities.} %\citep{HS2003,bosch2018}.}

% These are the \texttt{mixcoatl} outputs for $f$, $\vec{\ell}$, and $I_{xx}$, $I_{xy}$, $I_{yy}$, respectively.

\subsection{Vibration Correction}\label{sec:vibration}
After completion of Run 5 data taking, \jef{systematic effects caused by the }vibration of the spot projector setup \jef{were} identified in the collected images. \jef{When the XY stage used to position the projector decelerated to a stop after a dither, a} vibration \jef{with a long settling time was induced.} As a result, the \jef{FWHM of the spots was systematically increased.} The effects were significant on a sub-percent level for the positions, sizes and shapes of the spots. 

The effects were removed by the following procedure. We assumed that the measurement residuals, e.g., second moments of brightness $\delta I_{ij}$, are to be zero on average. The measurement residuals for $k$-th exposure were fitted with a plane $a_kx+b_ky+c_k$ by varying $a_k, b_k, c_k$ for each exposure so that we minimize the summed square of the difference between the measurement residual and the plane model. If the effect was constant across the exposure subtracting off $c_k$ should be enough. However, the subtraction of a plane from each exposure was needed empirically. The source could be the combination of the effects of vibrations and the tilt of the projector with respect to the focal plane that are different in different dithered exposures. The operation was applied for the $\delta \ell$, $\delta I_{ij}$, and $\delta f$ quantities exposure-by-exposure. In particular, this calibration had to be done in the second moments of brightness rather than the final products, such as ellipticities and shear, because they are not linear quantities under this operation.    

\section{Distortions of Photometry, Centroid, PSF Size and Shape}\label{sec:distortions}
In this section, we report our measurements of the intrinsic sensor distortion in photometry, centroid, and PSF shape and size for six LSSTCam sensors. Then we describe the most important distortions revealed by the residual maps of the two CCD designs, ITL and e2v. The physical nature of the effects is discussed in Sections \ref{sec:tree-ring}, \ref{sec:other-effects}. 

\subsection{Effects of Sensor Anomalies on measurements}\label{sec:report}
\begin{figure*}[!htb]
    \centering
    \includegraphics[page=4, width=0.90\textwidth]{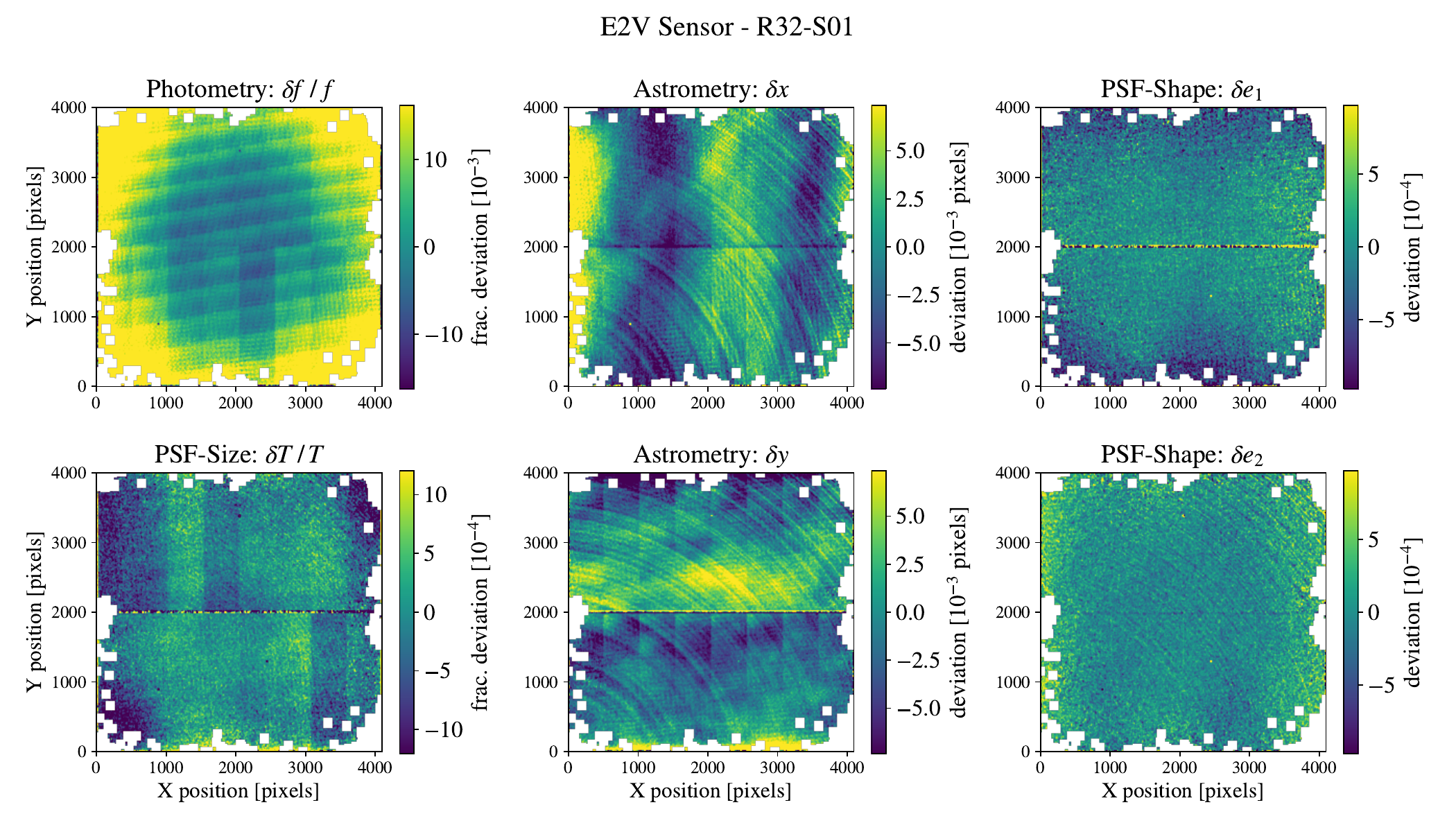} 
    \includegraphics[page=2, width=0.90\textwidth]{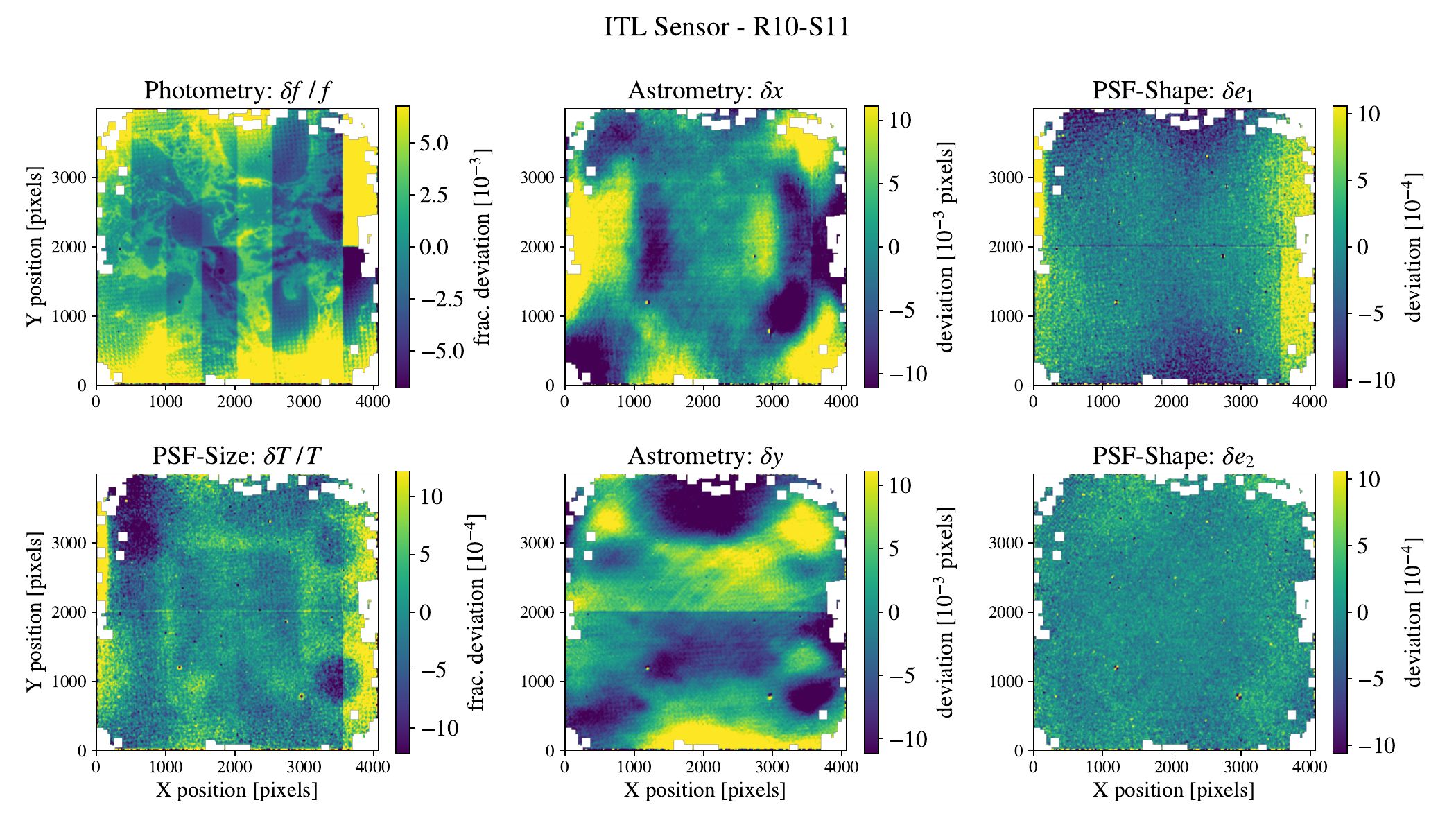}
    \caption{Maps of deviations of measurement quantities for two LSSTCam sensors, top: E2V R32-S01 and bottom: ITL R10-S11. The measured quantities are flux, PSF size, centroid, and shape. 
    Shifts of centroid and shape shown in $x$ and $y$ components separately. 
    The photometric map corresponds to a flat image for the sensor but made by stitching measurements of spots. Quantum efficiency variations across amplifiers are evident features in the these maps. The structures in the PSF size, position, and shape residual maps are caused by pixel-area variation effects. Some striking examples are the tree-ring pattern and the amplifier boundary effects. 
    % Note: the lower-right corner of R03\_S12 is an image artifact, the spot images in this region are likely affected by stray light reflections from the side of the cryostat window. 
    }\label{fig:distortion_map}
\end{figure*}

We show the intrinsic sensor pixel response distortion maps in  \autoref{fig:distortion_map}, which were created using calibrated flux, centroid, PSF size, and shape deviations measurements to reveal sensor anomalies as described in \autoref{sec:residuals}. 

We used a large collection of star-like exposures to obtain highly precise measurements. The faint sources were masked by selecting only the top 80\%-th percentile of flux (see \autoref{fig:spotgrid_illustration}). In addition, we binned the deviation measurements by pixel location and used super-pixels of size 10 $\times$ 10 pixels to create the deviation maps, which were then stretched back to their original size.

For instance, in Run 5, which consisted of 2000 exposures, we achieved an \je{RMS error on the astrometric shift maps} of $10^{-4}$ pixel (physical size $= 1$ nm) , and $10^{-5}$ on PSF-size/shape maps. Each exposure contained up to 2401 sources, resulting in an average of 10 sources per super-pixel. 

The deviation maps for an E2V and an ITL sensor are presented in \autoref{fig:distortion_map}, with six different measurements shown for each sensor. Additional deviation maps for other sensors can be found in \autoref{appendix:A}. To relate to the electric fields in the serial and parallel transfer directions of a sensor, we present the deviation maps separately for the x and y components. A brief description of each deviation map is provided.

\textbf{Photometry deviations:} This map corresponds to a \je{``star flat''} image but is made by stitching the deviations of flux measurements from the calibrated artificial stars. The e2v and ITL sensors display two distinct features in the maps: a rectangular shape associated with the 16 CCD segments, and irregular patterns resembling ``coffee stains" in the ITL sensors and an ``annealing pattern"\je{, the diagonal stripes,} in the e2v CCDs. \je{The irregular patterns are created by the surface finish in the silicon manufacturing process.} In particular the e2v sensor exhibits a radial gradient from the center to the corner, \jef{and} bright corners \jef{are} also observed in the ITL sensor. These features \jef{are} likely to be a large-scale residual pattern \jef{from} our spot-projector dithers. Some amplifiers have different contrasts showing distinct mean flux levels. These amplifier variations are due to the uncertainties on the gain estimation. These features are discussed further in  \autoref{sec:QE}. \jef{The deviations presented here can be corrected in the on-sky survey by a flat-field correction because they are quantum efficiency variations, not pixel-area distortions. }

\textbf{PSF size deviations:} The maps on the lower left in each group present the measured variations in the size of the PSF. The variations are at the sub-percent ($\times 10^{-3}$) level. They are related to effects that increase or decrease the size of a point source. We do see a circular ripple pattern centered on the outside of the sensor.  We discuss this effect in \autoref{fig:distortion_map}.
An unexpected structured variation in PSF size is also observed on the ITL sensors, with circles in three corners and a square in the middle. This pattern matches the hardware CCD frame that holds the detector. We investigate this feature further in \autoref{sec:CCD-frame}. \je{It is important to note we do not see \jef{effects associated with} the annealing pattern for \jef{the e2v sensor or associated with the coffee stain feature for the ITL sensor}, which implies these features are not pixel-area distortions.}

\textbf{Centroid deviations:} The two maps in the middle column show the $x$ and $y$ centroid deviations. The measurement displacement from the calibrated source is of order $0.01 \, \text{ pixel} = 100 \, {\rm nm} = 2$\,mas on the LSST focal plane. The circular pattern of the tree-ring effect is the most noticeable feature in the centroid residual maps. We analyze this effect further in \autoref{sec:tree-ring}. In the ITL sensors, we do not see features related to the CCD shapes other than the mid-line break on the $\delta y$ map. In contrast, a strong mid-line signal is evident for e2v along with strong contrast between the CCD segments' edges in both directions ($x$ and $y$). Beyond those effects, we see irregular spatial feature variations in both CCDs.

\textbf{PSF shape deviations:} \jef{The two maps in the right-hand column show measured ellipticity variations from the calibrated spots. 
In R32-S01, three features are induced by 1) tree rings, 2) something else that causes a global variation, and 3) noisy regions where the population of spots is insufficient. General trends that $e_1$ traces the tree-rings component along the axes ($X \text{ and } Y$) and $e_2$ traces the component along the diagonal component $X=Y$ (i.e., $45^\circ$) are apparent; see the explanation in \autoref{sec:tree-ring-coord}. The global variation and the noisy region make trends less striking, especially at the CCD edges where the number of stacked artificial stars is lower. }

% \je{the e2v R32-S01 PSF shape residual map has tree-ring-related features similar to those of the centroid residual map. In particular, \jef{the $e_1$ component should vanish along the CCD diagonal ($X=Y$)} because of the coordinate transformation to polar coordinates; see the explanation in \autoref{sec:tree-ring-coord}}. In addition to tree-rings, the mid-line break, and some global variations are present. In \autoref{sec:boundaries}, we investigate further the distortions related to the CCD segment boundaries. 

\je{A summary and a comparison of the size of these effects are presented in \autoref{tab:signal_level}. The values in the table are the maximum absolute deviation values measured in the six CCDs examined in this study. Despite the small numbers, these features should meet the very restrictive requirements set by the LSST science goals \jef{(summarized briefly in the table)}. Complete descriptions \jef{may be found} in the LSST Science Requirements Document \citep{Ivezic2011}, the LSST Science Book \citep{LSSTScienceBook2009} and \citep[][hereafter DESC2018]{DESC2018}.}:
% and links to technical papers and presentations at https://www.lsst.org/scientists.

\begin{table*}
\caption{Summary of the LSSTCam sensor effects limits on photometry (phot), centroid (center) and PSF-size and shape.}
\centering
\begin{tabular}{ccccccc}
    \toprule
    \hline
    Origin                                     & Name                                    & \multicolumn{4}{c}{Level} & CCD Type           \\ \cline{3-6}
                                               & & {\scriptsize phot. res [mmag]}& {\scriptsize centr. res [pixel]} & {\scriptsize psf-size frac. dev} & {\scriptsize psf-shape dev}& \\ \bottomrule 
    \multicolumn{1}{c|}{Scientific Goal}       & {\footnotesize Requirements}            & $ < 10.0 $& $ <5\times 10^{-2}$ & $< 1\times 10^{-3}$ &  $ < 1\times 10^{-3} $  & {\scriptsize Both}  \\
    \multicolumn{1}{c|}{}                      & {\footnotesize References} & \footnotesize{(a,b)} & \footnotesize{(a,b)} & \footnotesize{(c)} & \footnotesize{(a,c)}& \\ \bottomrule                                      
    \multicolumn{1}{c|}{}                      & {\footnotesize Mid-Line Break}          & --        & $ <4\times 10^{-2}$ & $                 $ &  $ <  4\times 10^{-3} $  & {\scriptsize Both}  \\
    \multicolumn{1}{c|}{Pixel-Area Varations}  & {\footnotesize Hardware Imprints}        & --        & $ <1\times 10^{-2}$ & $< 1\times 10^{-3}$ &  $        $  & {\scriptsize Both}  \\
    \multicolumn{1}{c|}{}                      & {\footnotesize Amplifier Boundaries}    & --        & $ <4\times 10^{-3}$ & $                 $ &  $ < 4\times 10^{-4} $  & {\scriptsize Both}   \\
    \multicolumn{1}{c|}{}                      & {\footnotesize Tree--Rings}             & --        & $ <4\times 10^{-3}$ & $< 5\times 10^{-4}$ &  $ < 5\times 10^{-4} $  & {\scriptsize Both} \\ \hline
    \multicolumn{1}{c|}{}                      & {\footnotesize Radial Gradient}          & $ < 25.0 $ & --               & --           & --       & {\scriptsize E2V}  \\
    \multicolumn{1}{c|}{Photometric Response}  & {\footnotesize Residual Impurities}      & $ < 4.0 $ & --               & --           & --       & {\scriptsize Both} \\
    \multicolumn{1}{c|}{}                      & {\footnotesize Amplifiers Gain variation} & $ < 1.0 $ & --               & --           & --       & {\scriptsize Both} \\ \hline
    \bottomrule 
    \end{tabular}
    \label{tab:signal_level}
    \footnotesize{references: (a)\citet{LSSTScienceBook2009}; (b)\citet{ivezic19}; (c) \citet{2018arXiv180901669T}}
\end{table*}

\begin{itemize}
    \item psf-shape: uncertainties in the final LSST galaxy shear catalog are to be dominated by the statistical error \je{($\sim \delta e$)}, $10^{-3}$ \je{(dimensionless)}, which sets an upper limit on contributions from systematic errors (DESC2018).
    \item psf-size: the maximum acceptable PSF size bias $\delta T/T$ is $10^{-3}$ for the ten-year Rubin/LSST survey (DESC2018).
    \item photometric uncertainty: is expected to be below 10\,mmag \citep{ivezic19}.
    \item astrometric uncertainty: for a single image is expected to be 10 mas (0.05 pixel) in order to achieve proper motion accuracy of 0.2 mas/yr and parallax accuracy of 1.0 mas for the 10 year survey \citep{Ivezic2011}. 
\end{itemize} 

\je{The LSST requirements have been \jef{met with no corrections applied. Our findings show the small amplitude of} the sensor QE, amplifier boundary, mid-line break and tree-rings distortions\jef{, and highlight} the quality of the LSSTCam sensors. While these effects are small and the requirements are met, \jef{the effects are non-zero and can be mitigated with corrections at the image analysis stage.} In the following sections, we will analyze these features comprehensively, present our physical interpretations, and discuss their impact on the LSST survey. }

% \je{While most of the features presented in \autoref{tab:signal_level} met the requirements it's important to discuss their impact on the LSST survey and possible corrections. In the following sections, we present our physical interpretation of these features and the discussion of these topics.}

\section{Effects of edges, amplifier boundaries and mid-line break}\label{sec:boundaries}
This section closely examines the anomalies at the CCD sensors amplifier segments and edges. 

\begin{figure*}[!htb]
    \centering
    \includegraphics[page=4, width=0.95\textwidth]{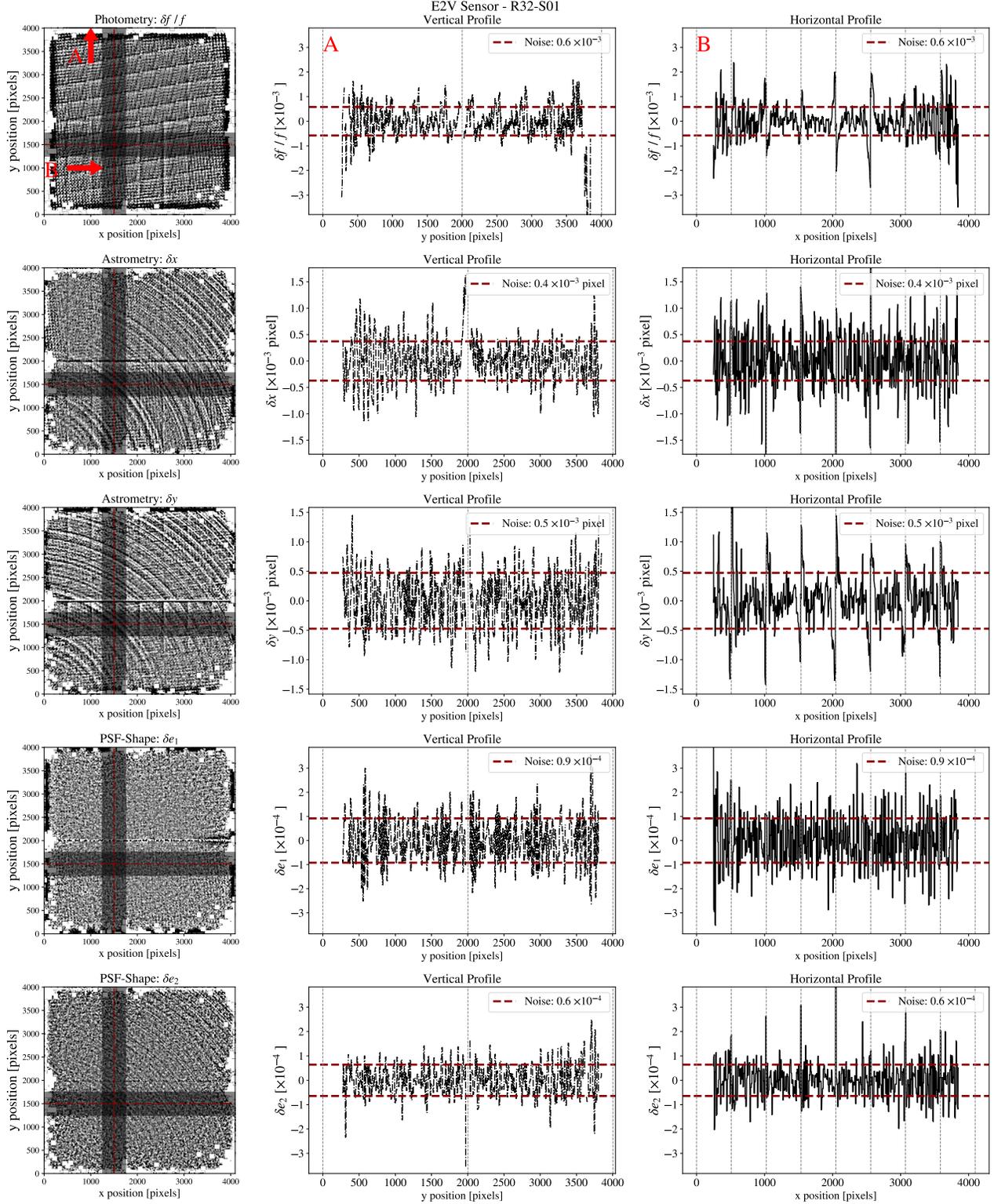}
    \caption{Distortion maps (left) and signal distortion profile across the horizontal (middle) and vertical (right) directions. A high-pass filter was applied to the images in the left-hand column to highlight the high-frequency distortions. The profiles are the map signal computed over the regions of the gray bands; the red dashed lines indicate the error level (rms). In the right, the $e_1$ and $\delta x$ horizontal profiles show the effects at the amplifier boundaries (gray dashed lines). In the middle column, the mid-line break is the most noticeable feature, especially for $\delta y$, with a signal level $100 \times $ higher than the noise. }
    \label{fig:signal_profiles}
\end{figure*}

The deviation maps have global features that can mask effects at the edges, amplifier boundaries, and the mid-line break. We remove such features by applying a high pass filter to the maps of variations with pixel periods higher than 250 pixels. \autoref{fig:signal_profiles} presents the cleaned residual map of the e2v R32-S01 sensor on the left. We compute the signal profile in the $x$ and $y$ directions for bands of 500 pixels width as indicated in the left-hand plot. The middle and the right-hand columns show the resulting vertical and the horizontal signal profiles, respectively. In the signal profile plots, the amplifier boundaries and the mid-line break are represented by grey dashed lines. We define the signal noise (red dashed lines) as the standard deviation of the signal.

The e2v sensor exhibits significant distortions in the vertical profiles, particularly at the mid-line break; \je{we masked the mid-line break for visualization purpose}. This feature is evident in all residual maps \autoref{fig:distortion_map}, the peak is about 100 times greater than the noise for the y-component maps. Away from the mid-line, we see other patterns in the flux profile, for instance, \jef{the residual surface patterns (annealing and coffee stains)} and increased distortion at the edges. \je{The surface features for e2v and ITL sensors are QE deviations since \jef{they are not found in} centroid or PSF measurements. }

% The photometric deviations have a pattern of distortions at the between  with amplitude on the order $10^{-3}$. 

%
\je{The mid-line break \jef{region has} the only distortions that surpasses the LSST survey requirements \citep{LSSTScienceBook2009}. The effect on the PSF measurements could \jef{be statistically significant;} in such case, corrections are needed. The simplest approach is to mask the region as we did in \autoref{fig:signal_profiles}, albeit at the cost of losing approximately $\Delta y = 20$ rows which corresponds to 0.5\% of the sensor area. An alternative solution is to flag the sources detected within these areas as lower quality. An elaborate approach \jef{would be} to model the distortion signal with a functional form, \jef{under the assumption that} the effect is static. This \jef{approach} is somewhat similar to the brighter-fatter effect correction \jef{in that regard} \citep[e.g.,][]{Lage2017}. Further assessment of the feasibility of such corrections is needed. For the amplifier boundaries, similar solutions \jef{as }for the mid-line break could be implemented. 
}

\section{Tree-rings}\label{sec:tree-ring}
In this section we focus on the signatures of tree-rings. We first present the relation between the flat-field distortion due to tree rings and centroid and PSF shape and size changes. Then, we describe our method to measure the tangentially averaged effects in the polar coordinate system.  Finally, we present the tree-rings oscillatory radial distortions and compare them with the flat-field signal.

\subsection{Pixel Area Variation}\label{sec:tree-ring-eqn}
In \citet{plazas14a,plazas14b}, the tree-ring effect in the DECam camera sensors is interpreted as effective changes of pixel area caused by a lateral electric field. Any quantities dependent on the pixel area are affected by this distortion. Here we follow the same approach to interpret the tree-ring effect in the LSST sensors in terms of pixel area changes.

The centroid shifts are modeled as the displacement $d(r)$ of the centroid of photons incident at a radial distance $r$ from the inferred center of the tree-rings compared to the actual radial displacement from the center $r_{0} = r - d(r)$. Then, the corresponding area distortion, $w(r)$, can be calculated as the Jacobian determinant of the coordinate transformation $r\rightarrow r_{0}$ \citep{plazas14a,plazas14b,Okura_2015}:

\begin{equation}
1 + w(r) = \left| \frac{r_0 dr_0 d\theta}{rdrd\theta} \right| \approx \left | \frac{(r-d(r))(dr-\partial_r d(r) dr)}{rdr} \right |\; ,
\end{equation}

At the first order one can show:
\begin{equation}
    w(r) = -\frac{\partial d(r)}{\partial r} - \frac{d(r)}{r} \approx  -\frac{\partial d(r)}{\partial r} \; .
    \label{eq:treering_integral}
\end{equation}
\je{The second term is negligible for $r > 10^{3}$ pixels where the LSST tree-rings signal is nonzero, given that $d(r)$ is on the order of ten times larger than $\partial_r d(r)$.} Note that we follow the definition  of \citet{Okura_2015} for the distortion of $d(r)$. As a result, equation \autoref{eq:treering_integral} has sign opposite to the convention in \citet{plazas14a}. 

% \johnny{need some rephrasing}
Further, we can define magnification in size $T^{1/2}$ and change in ellipticity $\delta e$ and relate them to the area perturbation $w(r)$. The perturbation on $T^{1/2}$ can be written as \citep{Okura_2015}:

\begin{equation}
    \frac{(T+\delta T)^{1/2} -T^{1/2}}{T^{1/2}}  = -\frac{1}{2} \left(\frac{\partial d(r)}{\partial r}+ \frac{d(r)}{r} \right) = -\frac{1}{2} w(r) \; .
\end{equation}
For a first order expansion,
\begin{equation}
    \frac{(T+\delta T)^{1/2} -T^{1/2}}{T^{1/2}} \approx \frac{\delta T}{2T} \; .
\end{equation}

Thus, the fractional change in PSF size is equal to the opposite of the flat-field distortion:
\begin{equation}\label{eq:tree-ring-psf-size}
    \frac{\delta T}{T}= - w(r) \; ,
\end{equation}

Similarly, the change in shear due to the tree-rings $\gamma_{TR}$ \citep{Okura_2015} in conjunction with equation \ref{eq:treering_integral} can be shown to be:
\begin{equation}
\begin{aligned}
\gamma_{TR}(r) & = \frac{1}{2}\left(\frac{\partial d(r)}{\partial r}-\frac{d(r)}{r}\right)=-\frac{1}{2}\left(w(r)-2 \frac{d(r)}{r}\right) \\
& \approx -\frac{1}{2} w(r) \; .
\end{aligned}
\end{equation}

Our ellipticity definition, \autoref{eq:secondMoments-shape}, is a factor 2 times the shear \citep{Schneider2005}; thus we can write:
\begin{equation}\label{eq:tree-ring-psf-shape}
    \delta e_r \approx 2 \gamma_{TR} \approx -w(r)  \; .
\end{equation}
Equations \ref{eq:tree-ring-psf-size}, \ref{eq:tree-ring-psf-shape} equate the PSF size and shape distortions to the opposite of the flat-field distortion $w(r)$.

\subsection{Tree-Rings Coordinate System}\label{sec:tree-ring-coord}
The centroid and shape tree-rings distortions depend on the coordinate system. In contrast, the PSF-size and the flat-field distortions are invariant under a change of coordinate system. To measure the tree-ring distortion effects, we transform $(\ell_x,\ell_y)$ and  $(e_1, e_2)$ to polar coordinates by applying the rotation matrix. The main difference between centroid and shape is the rotation matrix. Ellipticities are pseudo-vectors, thus, their transformation is:
 
\begin{equation}\label{eq:tree-ring-tangential-shear}
\left(\begin{array}{c}
e_r \\
e_\theta
\end{array}\right)=\left(\begin{array}{cc}
\cos 2\phi & \sin 2\phi \\
\sin 2\phi & -\cos 2\phi 
\end{array}\right)\left(\begin{array}{c}
e_1 \\
e_2
\end{array}\right)  \; ,
\end{equation}

where $\phi=\tan^{-1} \left(\frac{y-y_0} {x-x_0} \right)$. The center of this coordinate system $(x_0, y_0)$ is the center of the tree-rings circles, which is outside the CCD. The factor $2\phi$ is a  consequence of the invariance of ellipticities against $180^{\circ} $ rotations. \jef{For this reason, the tree-rings signal in the $e_1$ component is always less than for the $e_2$ component at the CCD diagonal $X=Y$ (i.e., $45^\circ$),  since the $ \cos(2\phi)$ term is zero for $\phi \approx 45^\circ \text{ or } 135^\circ$. One can visually confirm that the tree-rings signal vanishes at the diagonal in the $e_1$ PSF-shape maps (see \autoref{fig:distortion_map}). In contrast, this signal is higher along the $X$ and $Y$ axes as you can see in the maps with large tree-rings amplitude, e.g. R24-S11 and R02-S02 (Figures \ref{fig:fig1_a}, \ref{map:r22s11_r02s02}). However, for the sensors with low tree-ring amplitudes the noise at the CCD edges hides the rings features. }

\subsection{Algorithm}\label{sec:tree-ring-methods}
As we saw in the previous sections, our residual maps have the tree-rings and other features, such as global variations or structures associated with the amplifier boundaries. Therefore, to determine the components due to tree-rings, we perform image processing described in the following steps: 

\begin{enumerate}
    \item Image pre-processing: We clean the residual images. As other effects impact the residual maps at larger angular scales, we apply a high-pass filter to highlight the tree-ring effects. First, we binned the maps by 8$\times$8 pixels to increase the signal-to-noise ratio of the features and apply a high-pass filter to remove the global variation with pixel frequency higher than 250 pixel, as described in \citet{Park_2020}. During this process, we also identify bad pixels and mask them. Finally, the image is stretched to the original size.  
    \item Polar Transformation: We convert the original rectangular CCD pixel image (x,y)  to the polar coordinate system with respect to the tree-rings center. We use (\texttt{warpPolar} from \textsc{opencv}) and the wafer center \citep{Park_2020}. In this operation, the output is an image in $(r,\theta)$ coordiinates where $r$ is the distance from the tree-rings center.
    In this step, we check whether the steps above are successful by examining the transformed image. If the tree-rings center is misidentified, the straight line of the tree-rings along with the $\theta$ direction would be tilted or disturbed. 
    \item Extracting the profile: We average the ($r, \theta$) image over $\theta$ to evaluate the profile in $r$. 

    % \item \je{The averaged 1D profile can still shows large spatial variations related to the CCD segments. We apply one more time the high-pass filter with the same cutout frequency of the 2D image.  }
    % This signal affect the mean values of the signal creating a large wavelength bias.
    % This procedure ensure that the tree-rings signal oscillates around the mean value.
\end{enumerate}

The tree-rings center can be approximated by the center of the silicon wafer in most cases \citep[see their figure 2.][]{Park_2017}. However, we noticed a center mismatch in some cases after visually inspecting the polar images. Since these mismatches affected the tree-rings signal, we measured the tree-rings center on flat-field images with a back-bias voltage equal to zero, as the signal is more prominent at this voltage setup \citep{Park_2017}. Following \citet{Park_2017,Park_2020} algorithm, we fit tree-rings center values using 30 flat-field images. This verification test showed that the differences between the silicon wafer center and the actual center have an r.m.s of $75$ pixel. For the cases where the signal was affected by the center mismatch we use the fitted values. 

\subsection{Measured Distortions Due to Tree Rings}

\begin{figure*}
    \centering
    \includegraphics[page=4, width=1.0\textwidth]{figure3.pdf}
    \includegraphics[page=1, width=1.0\textwidth]{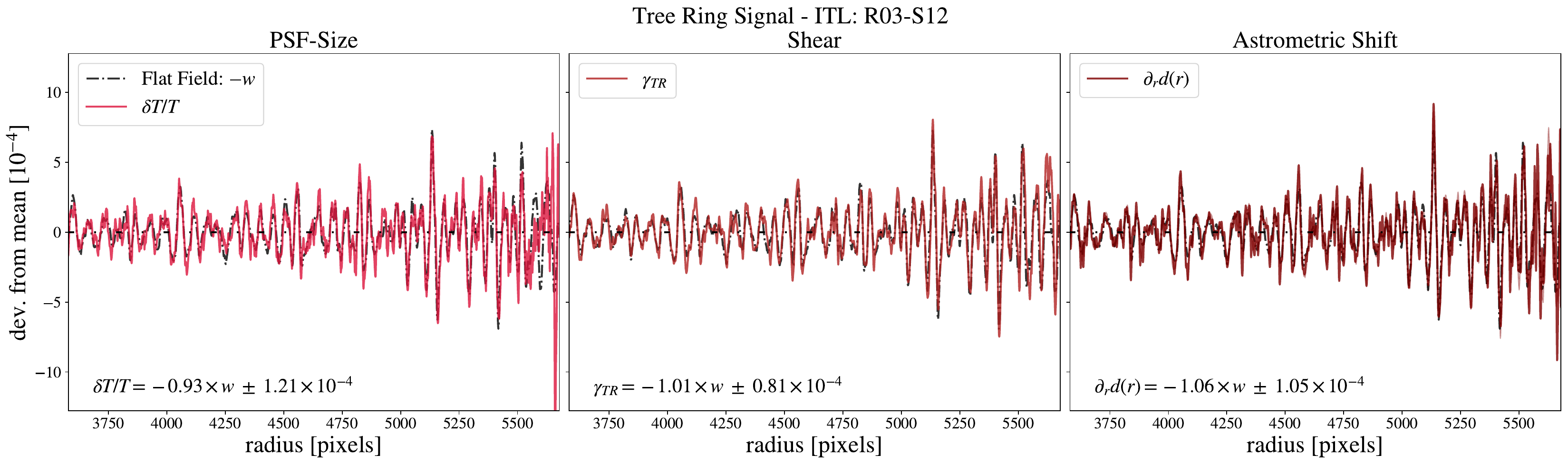}
    \caption{Distortions due to tree-rings for PSF size (left), shape (middle) and centroid (right) deviations as a function of distance from the the tree rings center for the sensors R32-S01 (top) and R03-S12 (bottom).  \je{The flat field distortion $w(r)$ (black dashed line) signal is correlated with the tree-rings signal (colored solid lines). In the lower-left corner \jef{of each panel,} the fitted linear relation between the two signals is displayed alongside the RMS error of the fit. The slope is close to unity, confirming the tree-rings effect. To highlight the oscillatory features of the signal, the radius range was limited to $2100$ pixels up to the maximum radius.} Note that the centroid shift is related to the derivative of the tree-rings distortions. }
    \label{fig:tree_ring_signal_panel}
\end{figure*}

Using the procedure described in \autoref{sec:tree-ring-methods}, we extracted the one-dimensional profiles of distortions due to tree-rings, finding amplitudes at the $10^{-4}$ level for centroid, PSF shape, and size distortions. 

\je{\autoref{fig:tree_ring_signal_panel} shows the radial profiles of the oscillating distortions. For the centroid shift, we show the derivative of the centroid shift with respect to the radius. In this case, we used a \texttt{savgol\_filter} with a window size of the typical tree-rings frequency, 72 pixels \citep{Park_2020}. }

As described in the \autoref{sec:tree-ring-eqn}, the tree-rings distortions of centroid and PSF shape and size are directly related to the distortions inferred from flat-field images; see, e.g., \autoref{eq:tree-ring-psf-shape}. For this reason, on each plot we overlay the distortions inferred from flats. Although measured distinctly, the distortions that we infer from tree-ring effects are very close to the direct measurement of the flat-field image distortions presented in \autoref{sec:distortions}. The amplitudes and the phases of the oscillations matches qualitatively the flat-field signal. 

To confirm the pixel-area distortions relations (\autoref{sec:tree-ring-eqn}), we fit the normalization factor from equations \ref{eq:treering_integral}, \ref{eq:tree-ring-psf-size}, \ref{eq:tree-ring-psf-shape}. The fitted relation results are presented in the lower left corner of each panel in \autoref{fig:tree_ring_signal_panel}. Overall, the fitted parameters are close to unity, validating the tree-ring effects impact on pixel-area quantities. The error of the fit indicated by the RMS error is on the order of $(5-10) \times 10^{-5} $, which is about the same noise we measured in the vertical and horizontal profile for PSF-shape (\autoref{fig:signal_profiles}). 

% The correlation coefficients close to unity (see \autoref{fig:tree_ring_signal_panel}) confirm the visual similitude and validate the relation between the tree-ring effects on pixel-area quantities as expressed by equations \ref{eq:treering_integral}, \ref{eq:tree-ring-psf-size}, \ref{eq:tree-ring-psf-shape}. Also, the amplitudes of the oscillating patterns increase with distance from the center as studied in \citep{Park_2017}. 

\subsection{Corrections: Tree-Rings}\label{sec:tree-ring-corrections}
The centroid displacements due to tree-ring effects are smaller than needed to satisfy the LSST science requirements on centroid \citep{2018arXiv180901669T}. Nevertheless, given the exquisite control on systematics demanded by LSST, the tree-ring effects could be corrected by measuring their radial profiles from flats (and photometry) and star flats (centroid; or using the formula \ref{eq:treering_integral} to extract the profiles of centroid shifts from flats). The profiles would be incorporated as templates in the centroid and photometric solution optimization during image reduction, with an amplitude parameter that depends on the filter band ( \citep[e.g.: ][]{plazas14a,plazas14b,bernstein17_astrometry,bernstein18_photometry,bernstein17_detendring}). 

\je{\jef{The approach of \citet{bernstein17_astrometry}} could be incorporated into the pipelines \citep{bosch2018, Bosch2019} for correcting centroid distortions. Although the DECam detectors show PSF size distortions similar to the tree-rings even after applying \jef{their} methodology \citep{Jarvis2021}, we believe that would not be the case here. \jef{For} DECam, the remaining PSF size distortions are likely due to charge diffusion. The LSSTCam PSF-size distortions follow a one-to-one relation with the flat-field signal (see \autoref{fig:tree_ring_signal_panel}), indicating that the main contributions are predominantly pixel-area distortions as demonstrated in \autoref{sec:tree-ring-eqn}. 
}

\section{Other effects}\label{sec:other-effects}
In addition to the tree-ring effects, we discuss a couple of other effects that we observed in \autoref{sec:boundaries}.

\subsection{Quantum Efficiency Variations}\label{sec:QE}
The patterns in the photometric residual maps for e2v and ITL are different. The detailed patterns can be interpreted as quantum efficiency variations that are caused by the back-side surface finish in the manufacturing process. The ITL pattern in the flux distortion map  (coffee stains) is due to a layer of non-stoichiometric oxidized silicon and cleaning residue of acid on the silicon surface right after etching, creating some non-uniformity in backside charging \citep{2020SPIE11454E..2DB}. Instead, the e2v sensors have a regular striped pattern in the flux distortion map which appears to be caused by the laser annealing process after the thinning process in the CCD fabrication \citep{Burke2004,RADEKA2006,Bender2014}. This residual surface effect is generally greater at shorter wavelengths since the blue photons are converted close to the CCD back-side. The amplitude of this effect should be at most $10^{-2}$ and almost undetectable in redder wavelengths than $\sim$500 nm from the verification tests on flat-fields \citep{Park_2017, Roodman18}.

The origin of the radially symmetric gradients in e2v maps is not clear. We do not see a similar pattern in regular flat images taken with a flat illuminator. The dithering of the spot projector might cause this pattern; however, the fact that the maps for ITL based on the same projector dithering pattern do not have the same radially symmetric pattern suggests otherwise. Further investigation using on-sky images will give more understanding of this effect.

\subsection{Gain Mis-Matches Between Segments}
% The \autoref{fig:distortion_map} shows different mean flux values across the different segments within $10^{-3}$ variation. These differences can be a result of the gain (e$^-$/ADU) measurement uncertainty. 

The \autoref{fig:distortion_map} shows different mean flux values across the different segments within $10^{-3}$ variation. These differences can be a result of the gain (e$^-$/ADU) measurement uncertainty. 
\je{The accuracy of our gain determination \jef{was of the order} $10^{-3}$, which is \jef{comparable to the level of the} discontinuity.
To mitigate the discontinuity, an adjustment using the imaging region \jef{could be used.} However, this is outside the scope of this paper.}
%For instance, the gains for each segment are derived from fitting photon transfer curves (PTC)\citep[see eqn. 20;][]{Astier2019}, including non linear terms which stems from the Bright Fatter effect. However, the fitting error is XX. \je{Ask Andres}

In addition, there are extreme examples, for instance the R03-S12 sensor, for which the amplifier at the bottom right shows a percent-level gain contrast. This segment \jef{was affected} by stray light reflections from the side of the cryostat window. As a result, the shadow made a significant impact on the gain determination. 

\subsection{Mid-line Break \& Amplifier Boundaries}\label{sec:amp-boundaries}
The mid-line break, which divides the top and bottom halves of the CCDs, is the most striking sensor feature, clearly visible in the map of both types of sensor \autoref{fig:distortion_map}.
Also evident are features associated with the readout amplifier boundaries, which divide the sensors vertically.
These effects are stronger for the e2v CCDs and can be up to $\times 10$ greater than for ITL sensors. This difference stems from design differences for the electronic readouts. For instance, e2v CCDs have a physical boundary that blocks electron flow between the upper and lower halves. On the other hand, ITL sensors have less-prominent mid-line effects because they have channel stops in the electronics readout similar to the amplifier boundary structures. The importance of the mid-line can vary between the sensors of the same design and can be greater than reported here.

\je{Another important spatial distortion reported \jef{for LSSTCam CCDs} is edge effects \citep{2018SPIE10709E..1LB}. The metallization around the edges of the CCDs is set at a positive potential that induces a lateral electric field shift \jef{extending} up to $\geq$ 10 pixels into the bulk. The footprint of the artificial stars is not extended enough to cover the CCD edges, so we cannot provide any guideline\jef{s about edge effects} from this work. However, CCD edge effects also should be taken into account in image reduction.}

% \je{TO BE REVIEWED}

% {\color{red} The photometric distortion maps show sharp contrasts at the segments' edges; see \autoref{fig:distortion_map}. The flux values might have been affected by the gain (e$^-$/ADU) measurement. Although during our imaging processing (see \autoref{sec:image_processing}) we accounted for the gain differences between CCD segments we still see a residual gain at a sub-percent level between the segments. As the PTC gain determination is sensitive to additional sources of variance \citep{stubbs14}, a possible explanation is that the distortions at amplifier boundaries might have caused the gain mismatch between segments. 

% In addition, there are extreme examples, for instance the R03\_S12 sensor, for which the amplifier at the bottom right shows a percent-level gain contrast. This segment suffered by stray light reflections from the side of the cryostat window. As a result, the shadow made a significant impact on the PTC determination. 

% These results show that the accuracy of the gain determined from PTC flat pair measurement that we currently achieve still results in a sub-percent discontinuities between amplifiers by the current implementation of the ISR. This suggests that further gain adjustment across amplifiers or more sophisticated PTC gain determination is needed. }
% %

\subsection{Imprints From Hardware Structure: CCD Frame}\label{sec:CCD-frame}
\begin{figure*}[!htb]
    \centering
    \includegraphics[width=0.85\textwidth,bb=0 250 1024 518]{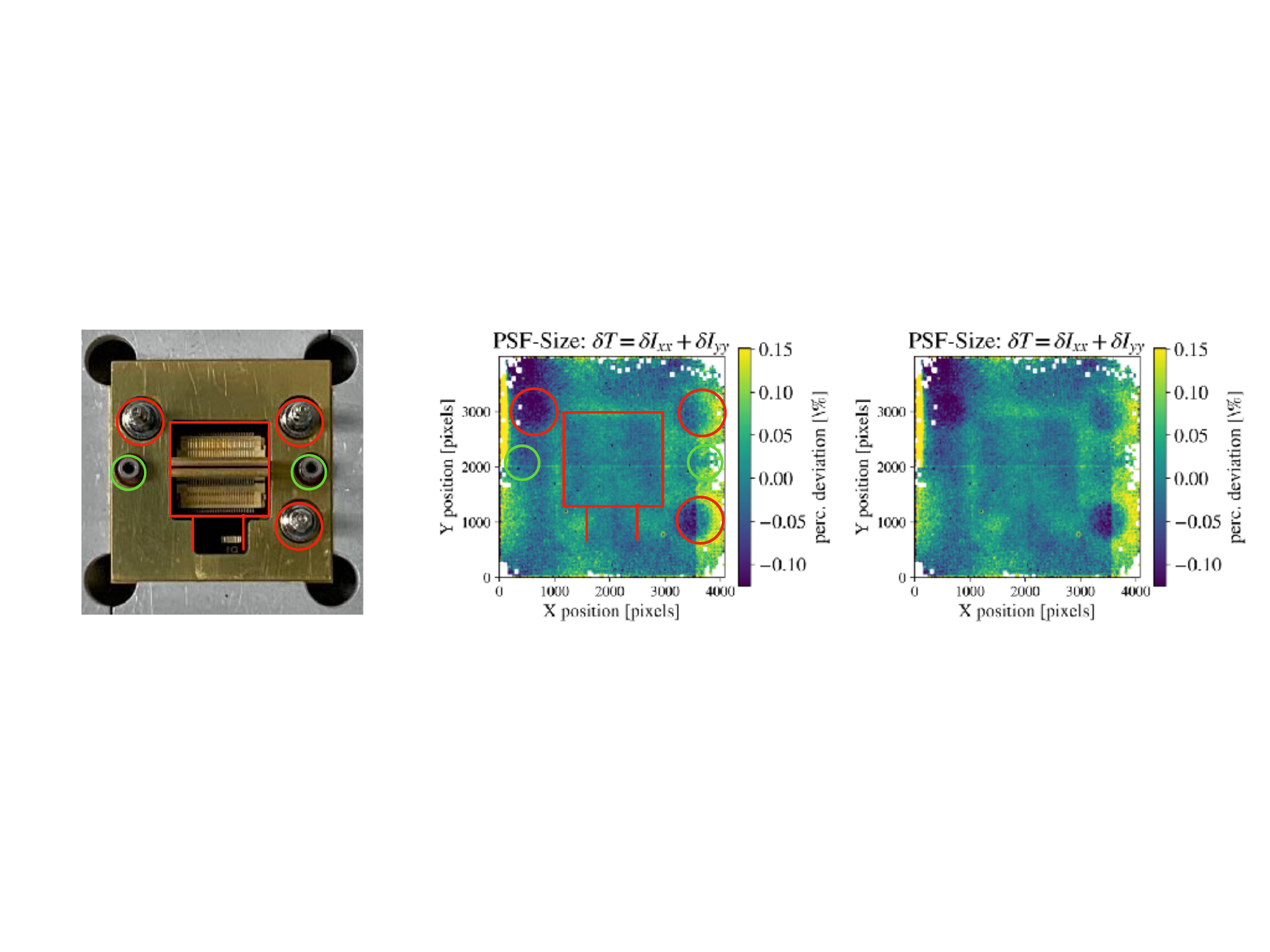}
    \caption{Left: The back structure of the ITL CCD (gold) is seen with two alignment pins (green) and three hold-downs (red). The connector to the flex cable is within the center box, where no gold metal support is present. Middle and right: PSF size residual map without and with a CCD frame drawing (red). The distortions of PSF size follow the shape of the CCD support frame.}
    \label{fig:hardware}
\end{figure*}

Multiple effects leave their imprints on the centroid and PSF-size maps of several sensors, including ITL's R03/S12, R10/S11, and e2v's R22/S11. In \autoref{fig:hardware}, the PSF size map of R10/S11 is shown alongside a photo of the ITL CCD support structure. Interestingly, the structure in the residual maps aligns with the location of the alignment pins and the hold-downs \citep[][]{Lesser_2017}. 

For the DECam CCDs, \citet[][see Figure 8]{bernstein17_astrometry} showed that the connectors impact their astrometry (centroid) when stacking all CCD images with different wavelengths (in the \textit{gri} bands). They concluded this is likely results from stresses induced in the CCD lattice by the connector or the hole in the mounting board. In addition, they also identified a wavelength-dependent feature related to the CCD frame metallic structure \citep{bernstein18_photometry}. The effect was strongest for the Y filter owing to the reflectivity of the metal structure, because a significant fraction of the infrared photons pass through the sensor. However, the ITL sensors for LSST have a highly IR-absorbing material ``lithoblack” deposited on the sensor wafer’s front side surface to prevent such an effect \citep{Lesser_2017}. In any case reflection of blue photons that would have passed through the CCD is implausible.

\je{This effect might potentially impact the survey, as affected regions are close to the 0.1\% threshold limit set by LSST requirements. For instance, other focal plane sensors not probed in this work may have greater distortions, which could leave imprints in astrometry and PSF-size data at the focal plane level. In contrast,  \citet{bernstein18_photometry} chose to leave \jef{the affected regions} uncorrected since they \jef{were not found to be associated with} significant effects. Further assessment of this systematic error on the LSSTCam focal plane CCDs should be performed. }
% At the quest of multiple LSST visits of a star, this signal could impact the LSST shear measurement as a general noise level that depends on the pixel position history.

\subsection{Artifacts}
The photometry distortion map of the sensor R22-S11 ( \autoref{map:r22s11_r02s02}; see top panel) has a noisy pattern. We do not have a clear explanation for this effect. It is likely caused by code failures related to determining the flux normalization since the other maps do not have corresponding effect. 

\je{Features caused by a fingerprint on the aperture window \jef{of the test apparatus were apparent} in few residual maps. The sensor ITL R02-S02 (\autoref{map:r22s11_r02s02}) has a visible shadow feature (roughly square) in photometry distortion maps at $x \approx 2500, y\approx 1500$.  A similar but less striking pattern appears in the astrometry and PSF size and shape measurements. For the ITL R03-S12 sensor (Figure \ref{fig:distortion_map}), a less striking feature can be seen at $x=2500, y=1500$ in the PSF shape and astrometry maps.}

\section{Discussion}
\subsection{Limitations}
\je{This work represents a pilot study to probe the \jef{impact of LSSTCam} sensor effects on the LSST survey. Although the findings and the data presented here are quantitative,
\jef{they are for only a small subset of the LSSTCam CCDs and} are not intended to be used for LSSTCam calibrations. The dataset for calibrations will be built during the first months of \jef{Rubin observatory} operations. Star flats should be created for the complete sets of sensors and filters. }

\je{Potential limitations of this study should \jef{be kept in mind.} The pre-processing described in Section \ref{sec:vibration} could potentially remove linearly varying features, or add any sensor systematics of larger size than the extent of the spots. To mitigate these \jef{considerations,} we introduced a dither\jef{ed pattern for the spot projector grid} and calculate\jef{d} the average values of measurements by averaging all 2000 dither\jef{ed images} spanning the CCD, shown in Figure \ref{fig:dithering_pattern}. The detail\jef{ed} features such as the annealing pattern, coffee stain, tree-rings are real, but global variations could potentially be \jef{artifacts} of this pre-processing. On-sky testing will answer the question.}

% In R22/S11 the photometry map in Figure \ref{map:r22s11_r02s02} there is unnatural discontinuity. This pattern could be caused by the projector related systematic issue. 

\je{Also, our study is limited to just a single wavelength \jef{range, around} 450\,nm. For the LSST survey the wavelength dependence of sensor anomalies should be evaluated, particularly their effects on PSF size and shape \citep{Meyers2015,Kamath2020}. Also, fringing effects in the y-band should be quantified and the proposed corrections for \jef{fringing} validated \citep{Guo2022}. }

\je{Finally, our study focused only on selected sensors. The sensor anomalies over the full focal plane as well as all other wavelengths will be investigated when the observatory collects on-sky data.}

% \subsection{Spot Data Usage}

%
\subsection{Tearing mitigation}\label{sec:tearing}
\jef{We observed tearing along the mid-line break in the Run 3 period.
This effect can be explained by the electric field distortions created at the boundaries between channels, caused by the non-uniform distribution of holes around the channel stop \citep[][see their Section 2.2.3]{Juramy_2020}. The cause \jef{of the tearing effect} has been attributed to the CCD readout procedure, in particular, the parallel clocking operation.  \citet{Juramy_2020} suggested mitigations via changing the readout voltage setup. Between Runs 3 and 5, significant efforts were implemented to mitigate this feature, including changing the readout voltages (see \autoref{table:operationvoltage})}
% Fortunately, only the tearing static version at the boundaries were observed in our study.

This change in operation voltage made a significant impact on the tearing features, mostly invisible to the eye. However, we did not find any obvious improvement or decrease of the effects at the amplifier boundaries and at the mid-line break. When we compare two set of profiles of R32-S01 (Run5) or R24-S11 (Run 5) and R22-S11 (Run 3) in the Appendix, the mid-line break and amplifier boundaries distortion are approximately the same. 

% This fact indicates that we need either 1) to mask out the adjacent pixels nearby the amplifier boundaries or the mid-line break, or 2) to develop further improvements to the \je{LSST instrument signature removal (known as ISR) procedure}. As the amplitude of theses segments effect is at the same level as the photometry residual maps, the electric field distortions is likely the cause. The same correction algorithm as for the tree-rings effect could be applied; evaluation is underway.

\subsection{Tree-Rings In Other CCD Devices}
The presence of tree-ring features is reported in other large CCDs cameras such as DECam \citep{Flaugher2015}, Hyper Suprime-Cam \citep{2014SPIE.9154E..1ZK}, and PanSTARRS GPC1 \citep{magnier18}. The relative amplitudes of the effect range from 0.1\% to 10\% in those systems.

\citet{plazas14b} reported the amplitude of tree-ring effects in DECam flat images is at the 1--10\% level and concluded their nature is pixel size variations by comparing their photometric and centroid deviations.  A similar effect has also been reported in the Hyper Suprime-Cam \citep{Kamata2014}.

In contrast, a unique effect, referred to as `charge diffusion', was identified in the GPC1 detectors \citep{magnier18}. Although this effect displays similarities to the tree-ring signal in terms of photometry and PSF size, it is actually driven by variations in the rate of vertical charge transportation. The primary outcome is a charge diffusion of variable length that predominantly affects the PSF size, but not the shape.

Later, the tree-rings effect on PSF-size was also seen in a few CCDs in DECam, after application of a correction for tree-ring distortion based on astrometric shifts \citep{Jarvis2021}. Their residual signal amplitude was much more prominent in the blue band, which indicates it occurs at the surface where light enters the CCD. Their could be interpreted as being due to charge diffusion. 

In this study, we compared $(dT/T)$, shape $\delta e$, and centroid shift $d(r)$ distortions and the flat signal $w(r)$. We fitted a linear relation between these signals and found \jef{near direct proportionality} as predicted by equations \ref{eq:treering_integral}, \ref{eq:tree-ring-psf-size}, \ref{eq:tree-ring-psf-shape}. Consistent with the findings of \citep{plazas14b}, we interpret the effects caused by tree-rings \jef{as being} due to shifts in parallel electric fields. Although charge diffusion variability effects can be present in the LSSTCam sensors the \jef{correlations} between  $dT/T$ and $w(r)$ \jef{did not significantly depart from unity.} Studies at redder wavelengths could potentially show different characteristics of the tree-rings signals on PSF-size due to diffusion effects.

\section{Conclusion}

This work represents the first probe of the impact of CCD anomalies on photometry, centroid and PSF measurements of the LSSTCam. We analyze the impacts using measurements artificial stars on a subsample of six LSSTCam sensors.
We classify a variety of distortions according to their source: CCD mid-line break, hardware imprints, amplifier boundaries and tree-rings. We report our main findings below.

\begin{itemize}
    \item \je{The centroid distortions are lower than the requirements on astrometric systematic errors of a single image \citep{ivezic19}}. The largest centroid distortions are due to the mid-line break, a design feature present only on the e2v sensors in LSSTCam. The ITL sensors have an unexpected hardware imprint from the metallic structure of the CCD frame \je{and the effect is not large than the mid-line}. The centroid distortions due to amplifier boundaries and tree-rings, are 10$\times$ smaller effects.
    
    % The centroid distortions are much less than the LSST Year 10 limits, below 0.04 pixel which is equivalent to 8\,mas. The greatest centroid distortions are due to the mid-line break, a design feature present only on the e2v sensors in LSSTCam. In contrast, the ITL sensors have an unexpected hardware imprint from the metallic structure of the CCD frame with a maximum distortion of 0.02 pixel. The centroid distortions due to amplifier boundaries and tree-rings, are 10$\times$ smaller.
    
    \item The shape distortions presented here are similar to the centroid ones. However, the amplitude of the mid-line break distortion in the shape measurements is slightly greater than the LSST requirement. Since the area affected is less than $0.5\%$ of the CCD area we suggest masking a region around the mid-line of e2v CCDs during image reduction. 
    
    \item The photometric distortions originate with spatial variation of quantum efficiency; their effect is not higher than 3\,mmag. The main features observed are due to gain variations between the sixteen CCD segments and global pattern features. The global features distortions have distinct visual appearance between the two CCD designs. For ITL sensors, the distortions have the appearance of `coffee stains' while for e2v sensors the appearance is due to laser annealing. The differences may be traced to differences in the manufacturing processes, in particular, the CCD back-side silicon treatment procedure.

    \item The tree-rings distortion effect measured for centroid shift is of the order of $10^{-4}$ pixel and $10^{-5}$ for PSF size and shape. These changes can be related to the distortions measured in flat-fields through the transformation defined in equations \ref{eq:treering_integral}, \ref{eq:tree-ring-psf-size}, \ref{eq:tree-ring-psf-shape}. Therefore, if necessary the effect can be corrected with the use of flat-field \je{template signals}. 

\end{itemize}

We find variations in distortion even among sensors from the same vendor. Further on-sky studies are needed to probe the variation of these effects for the 189 LSSTCam science sensors, as well as to study their dependence on wavelength. The laboratory study we present here  provides a foundation for understanding those effects in the entire focal plane using on-sky data.

\section{Acknowledgments}
We sincerely thank the anonymous referee for their thorough review of this manuscript and their invaluable feedback.
This material is based upon work supported in part by the National Science Foundation through Cooperative Agreement AST-1258333 and Cooperative Support Agreement AST-1202910 managed by the Association of Universities for Research in Astronomy (AURA), and the Department of Energy under Contract No. DE-AC02-76SF00515 with the SLAC National Accelerator Laboratory managed by Stanford University. Additional Rubin Observatory funding comes from private donations, grants to universities, and in-kind support from LSSTC Institutional Members. 

\bibliography{ref}

\appendix{}
\section{Operation voltages}\label{appendix:Operation}

The LSSTCam operation voltags setups using the during the data acquisitions are presented in \autoref{table:operationvoltage}.

\begin{table}[htb!]
\centering
\begin{tabular}{c | c | c | c | c } 
\hline                                                                
\hline
             & e2v (Run 3)  &    e2v (Run 5)  &     ITL (Run 3)  &   ITL (Run 5) \\
\hline                                                                
\hline                                                                
Output Drain [V] &  24.4        &    23.4         &     25.0         &   26.9       \\
Reset Drain [V]  &  12.7        &    11.6         &     13.0         &   13.0       \\
Guard Drain [V] &  26.0        &    26.0         &     20.0         &   20.0       \\
Output Gate [V] &  $-$2.2        &    $-$3.4         &     $-$2.0         &   $-$2.0       \\
Backbias    [V]   &  $-$50.0        &    $-$50.0         &     $-$50.0         &   $-$50.0       \\
\hline
Parallel Clock High [V]   &  3.4         &    3.3          &     2.0          &   2.0        \\
Parallel Clock Low [V]    &  $-$5.8        &    $-$6.0         &     $-$8.0         &   $-$8.0       \\
Serial Clock High  [V]  &  4.7         &    3.9          &     5.0          &   5.0        \\
Serial Clock Low  [V]   &  $-$4.2        &    $-$5.4         &     $-$5.0         &    $-$5.0       \\
Reset Gate High   [V]   &  6.4         &    6.1          &     8.0          &   8.0        \\
Reset Gate Low    [V]   &  $-$3.4        &    $-$4.0         &     $-$2.0         &   $-$2.0       \\
\hline
%csGate       &  0.0         &    1.0          &     1.0          &   1.0        \\
%AF1          &  0.0         &    0.0          &     0.0          &   0.0        \\
%Clamp        &  0.0         &    0.0          &     0.0          &   0.0        \\
Gain         &  0         &    0          &     0          &   0        \\
RC           &  3         &    14         &     3          &   14       \\
\hline
sequencer file & FP\_E2V\_2s\_ir2\_v2.seq &  FP\_E2V\_2s\_ir2\_v26.seq   &  FP\_ITL\_2s\_ir2\_v3.seq & FP\_ITL\_2s\_ir2\_v26.seq \\
\hline
\hline                                                                
\end{tabular}
\caption{Nominal operational parameters for both e2v and ITL sensors in the different run campaigns for which we acquired spot projector   data. See Figure 2 in \citet{2018SPIE10709E..2BS} for the schematic diagram of voltages. RC and Gain are configurable settings in  the readout electronics boards \citep{2014SPIE.9154E..1PJ}}
\label{table:operationvoltage}
\end{table}

\section{Supplemental Figures}\label{appendix:A}
We show the figures \ref{fig:distortion_map}, \ref{fig:signal_profiles}, and \ref{fig:tree_ring_signal_panel} for the other sensors studied: E2V - R24-S11 and R22-S11, and ITL - R10-S11 and R02-S02.

\begin{figure*}
    \centering
    \includegraphics[page=3, width=1.0\textwidth]{Figure1.pdf} 
    \includegraphics[page=1, width=1.0\textwidth]{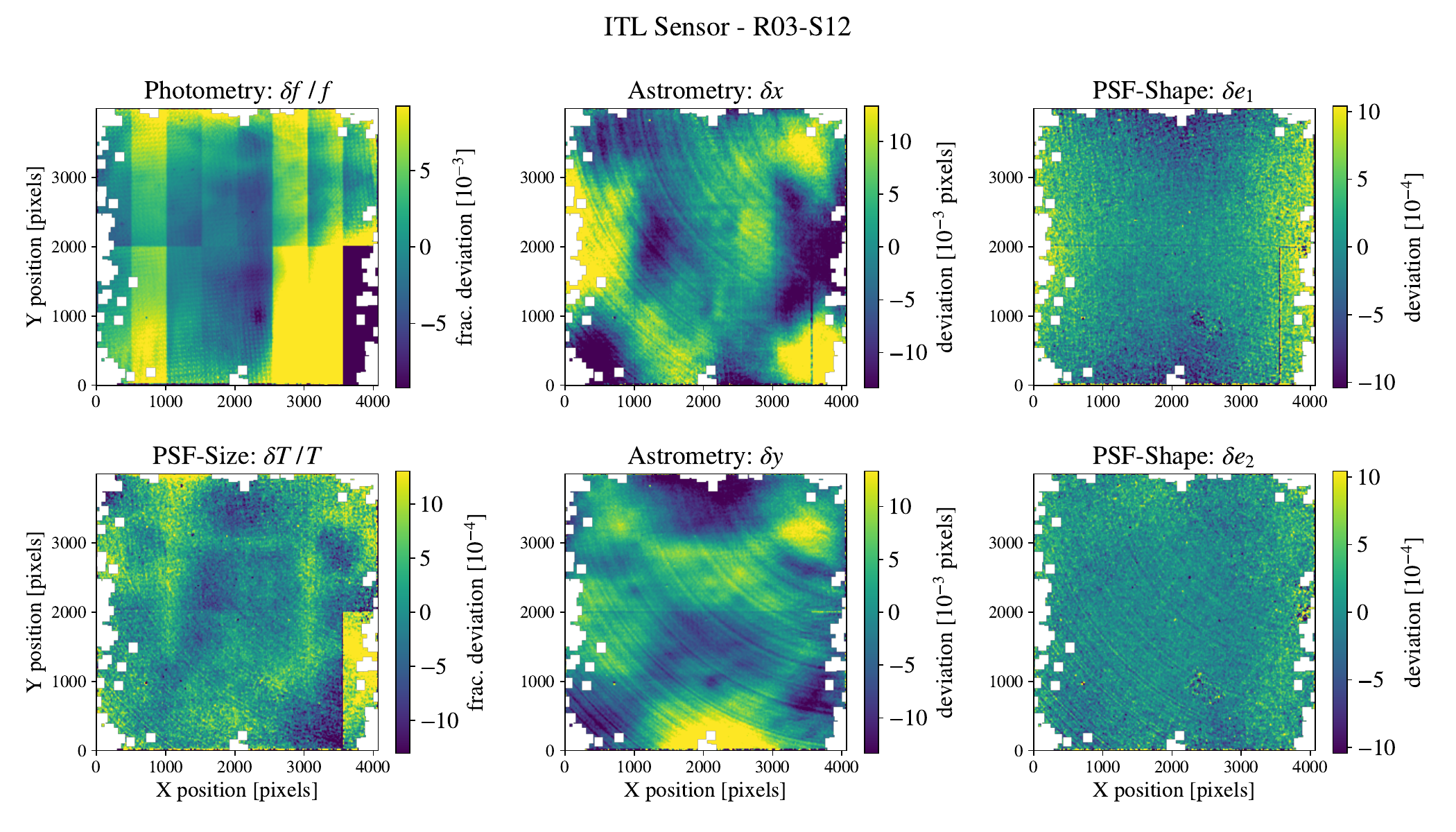}
    \caption{Same as \autoref{fig:distortion_map} caption. Note: the lower-right corner of R03-S12 is an image artifact, the spot images in this region are likely affected by stray light reflections from the side of the cryostat window. }
    \label{fig:fig1_a}
\end{figure*}

\begin{figure*}
    \centering
    \includegraphics[page=5, width=1.0\textwidth]{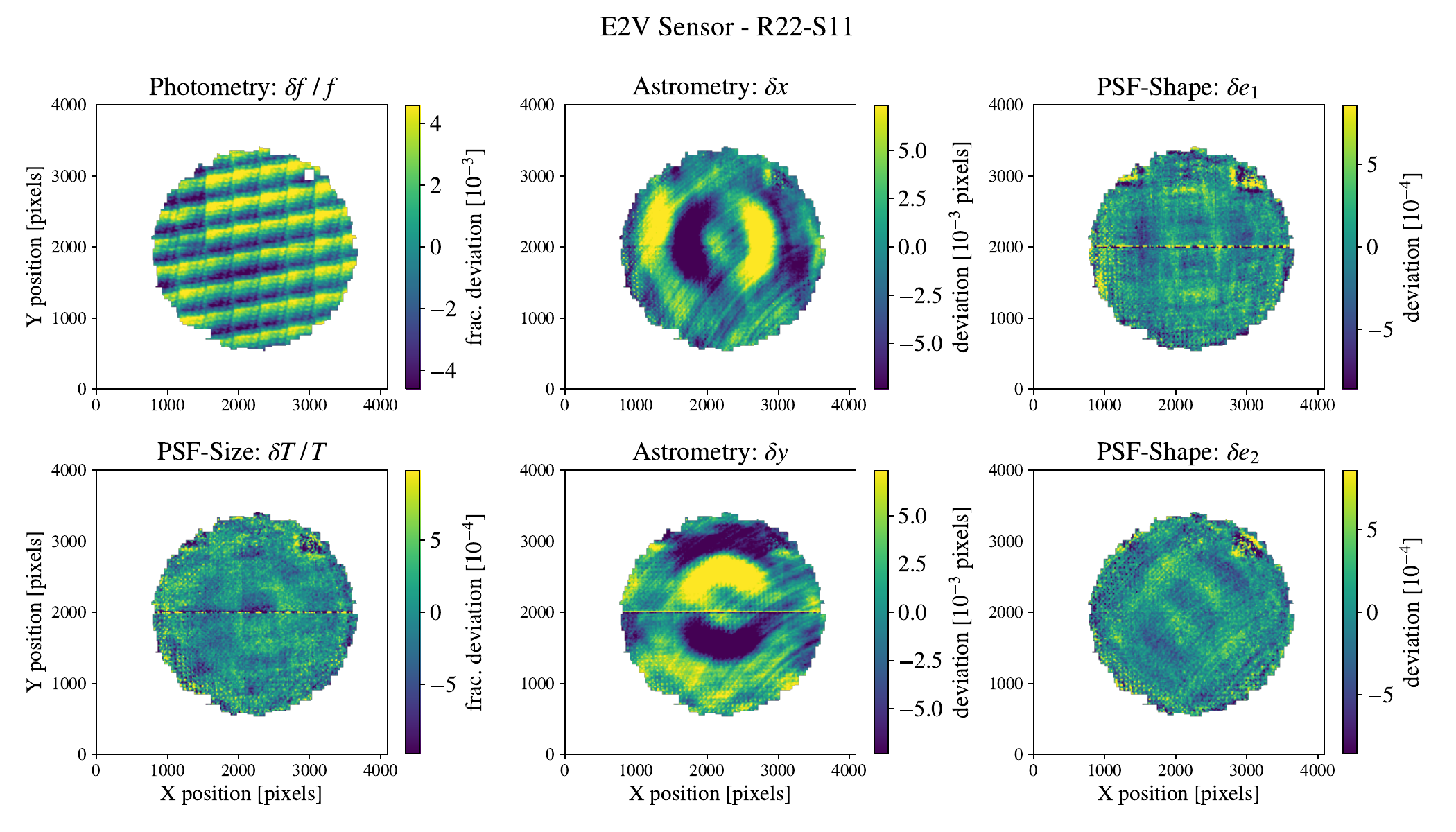}
    \includegraphics[page=6, width=1.0\textwidth]{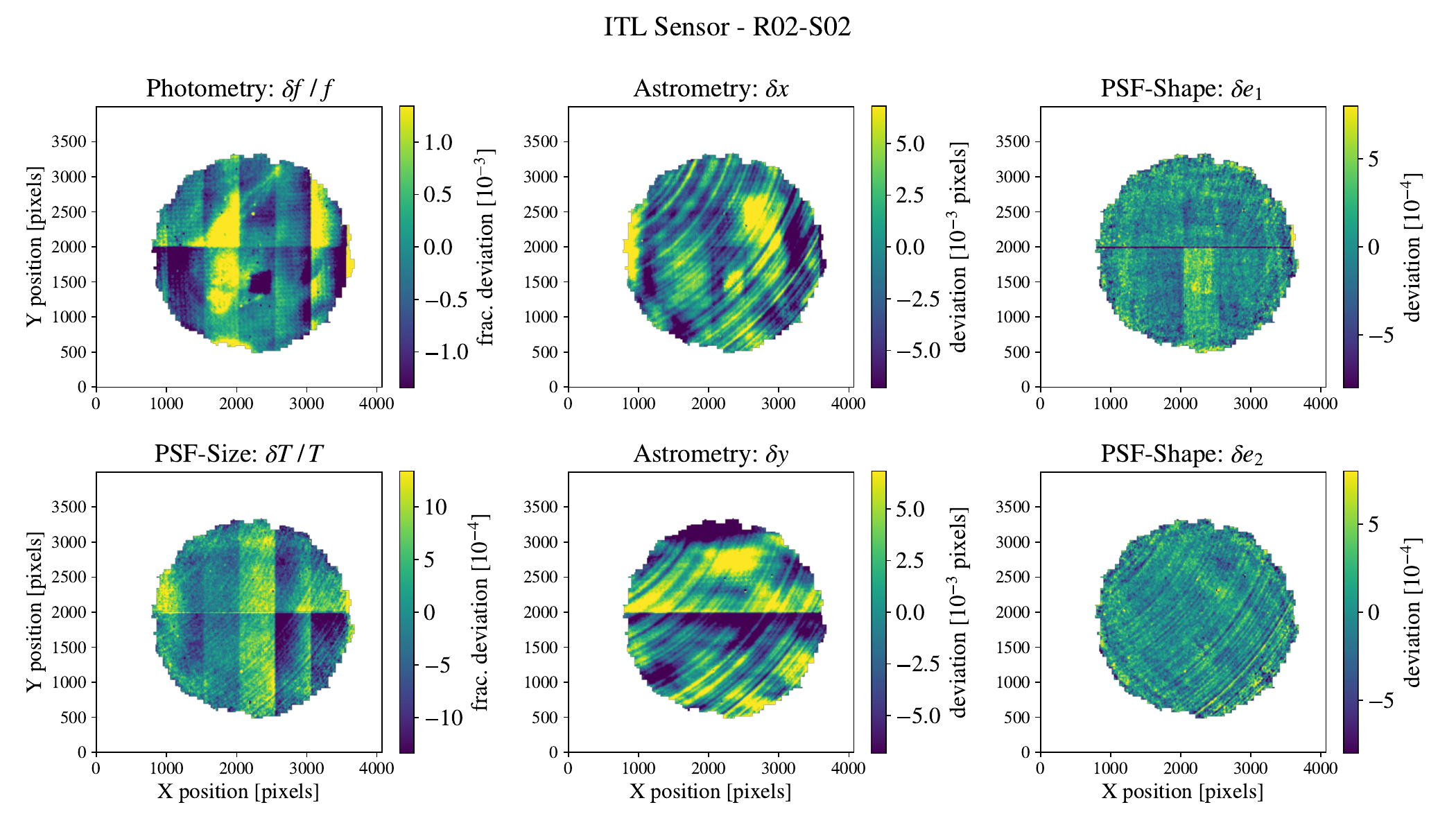}
    \caption{Same as \autoref{fig:distortion_map} caption. The data was collected during the Run 3 testing period. In this period the coverage of the fiducial spot projector position didn't extended to cover the entire CCD, which results in the circular shape map in the center of the CCD.}
    \label{map:r22s11_r02s02}
\end{figure*}

\begin{figure}[!htb]\label{fig:fig2_a}
    \centering
    \includegraphics[page=1, width=0.9\textwidth]{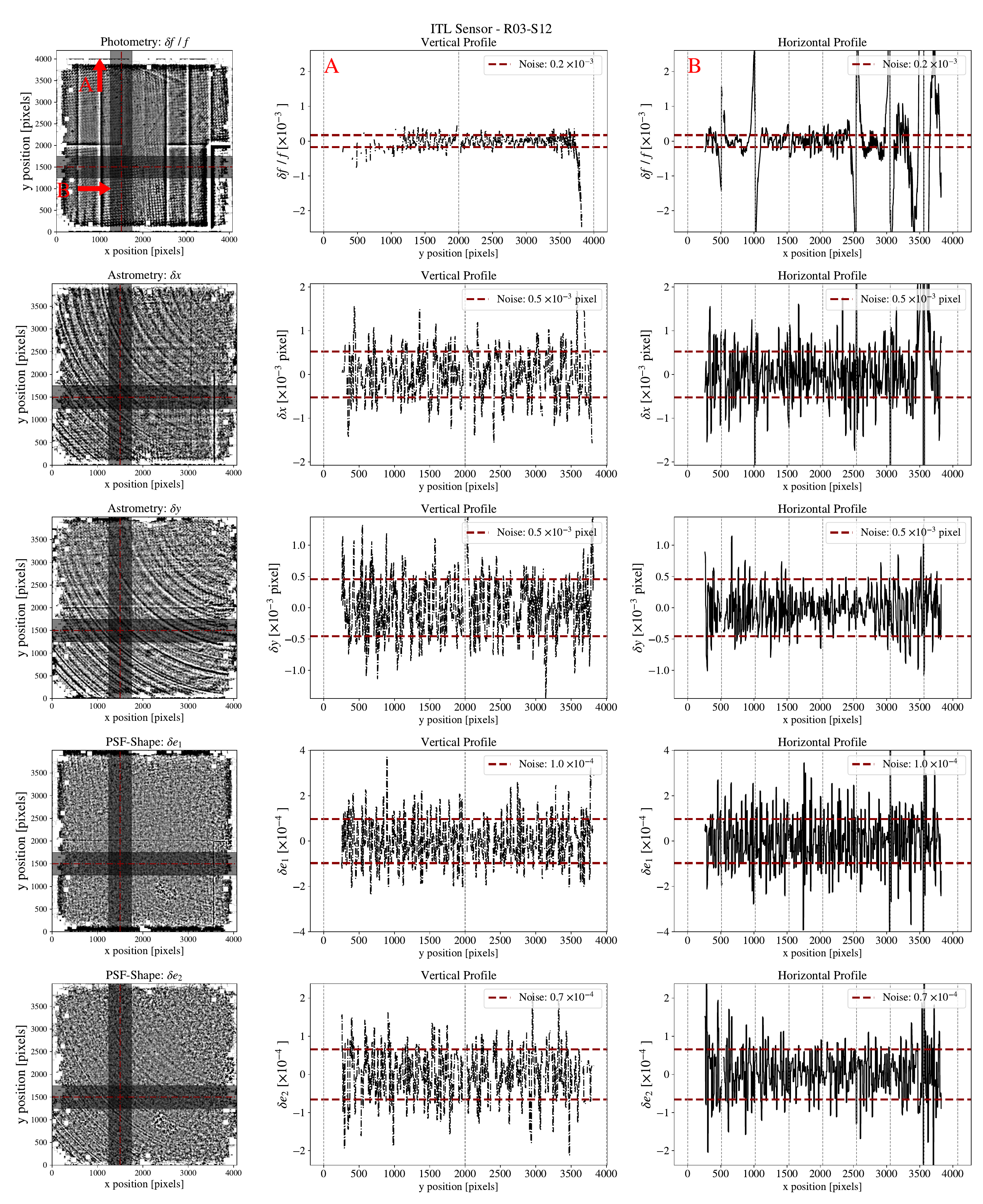}
    \caption{Same as \autoref{fig:signal_profiles} caption.}\label{profile:S03S12}
\end{figure}

\begin{figure}[!htb]\label{fig:fig2_b}
    \centering
    \includegraphics[page=2, width=0.9\textwidth]{figure2.pdf}
    \caption{Same as \autoref{fig:signal_profiles} caption.}
\end{figure}

\begin{figure}[!htb]\label{fig:fig2_c}
    \centering
    \includegraphics[page=3, width=0.9\textwidth]{figure2.pdf}
    \caption{Same as \autoref{fig:signal_profiles} caption.}
\end{figure}

\begin{figure}[!htb]\label{fig:fig2_d}
    \centering
    \includegraphics[page=5, width=0.9\textwidth]{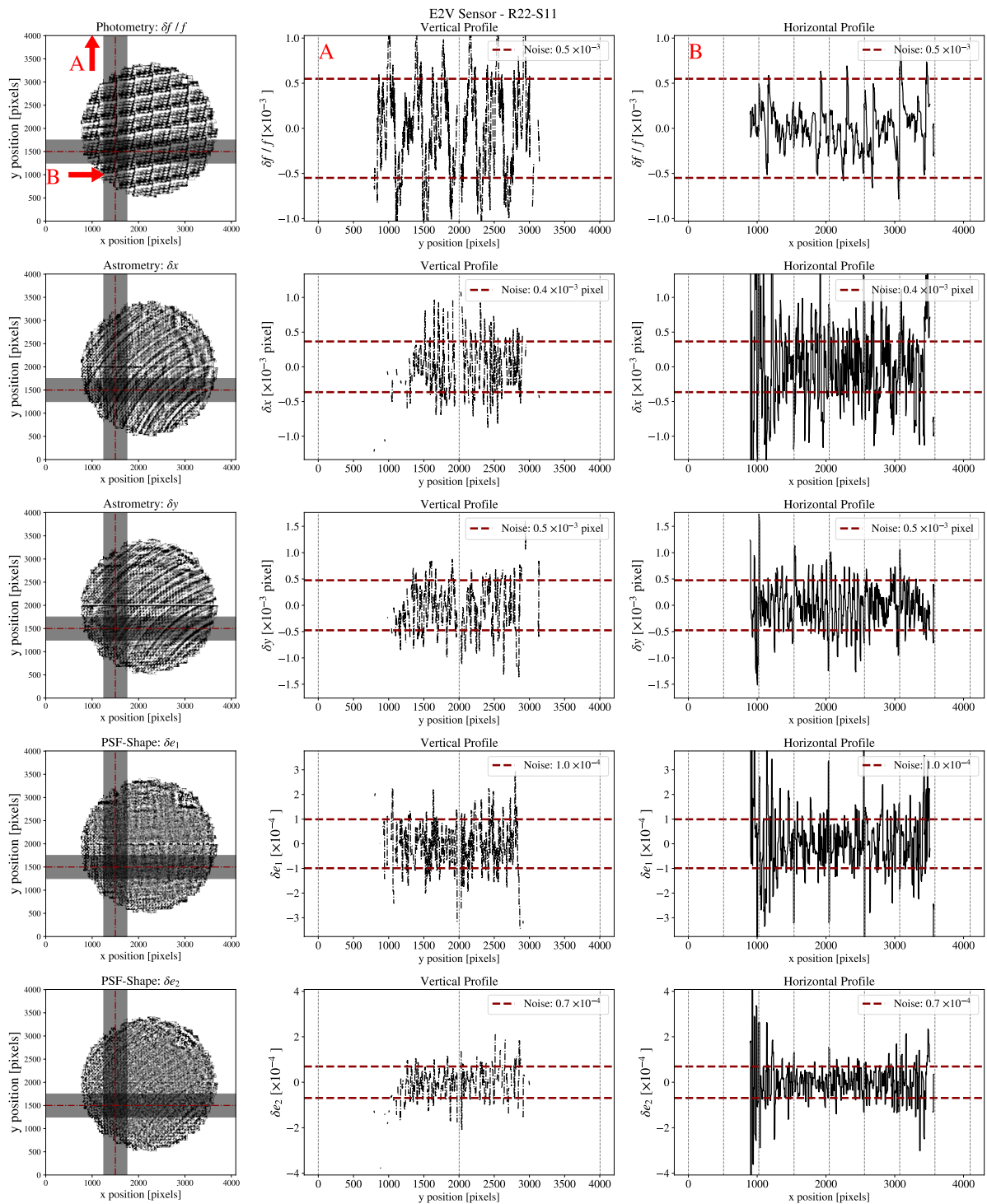}
    \caption{Same as \autoref{fig:signal_profiles} caption. The data was collected during the Run 3 period.}
\end{figure}

\begin{figure}[!htb]\label{fig:fig2_e}
    \centering
    \includegraphics[page=6, width=0.9\textwidth]{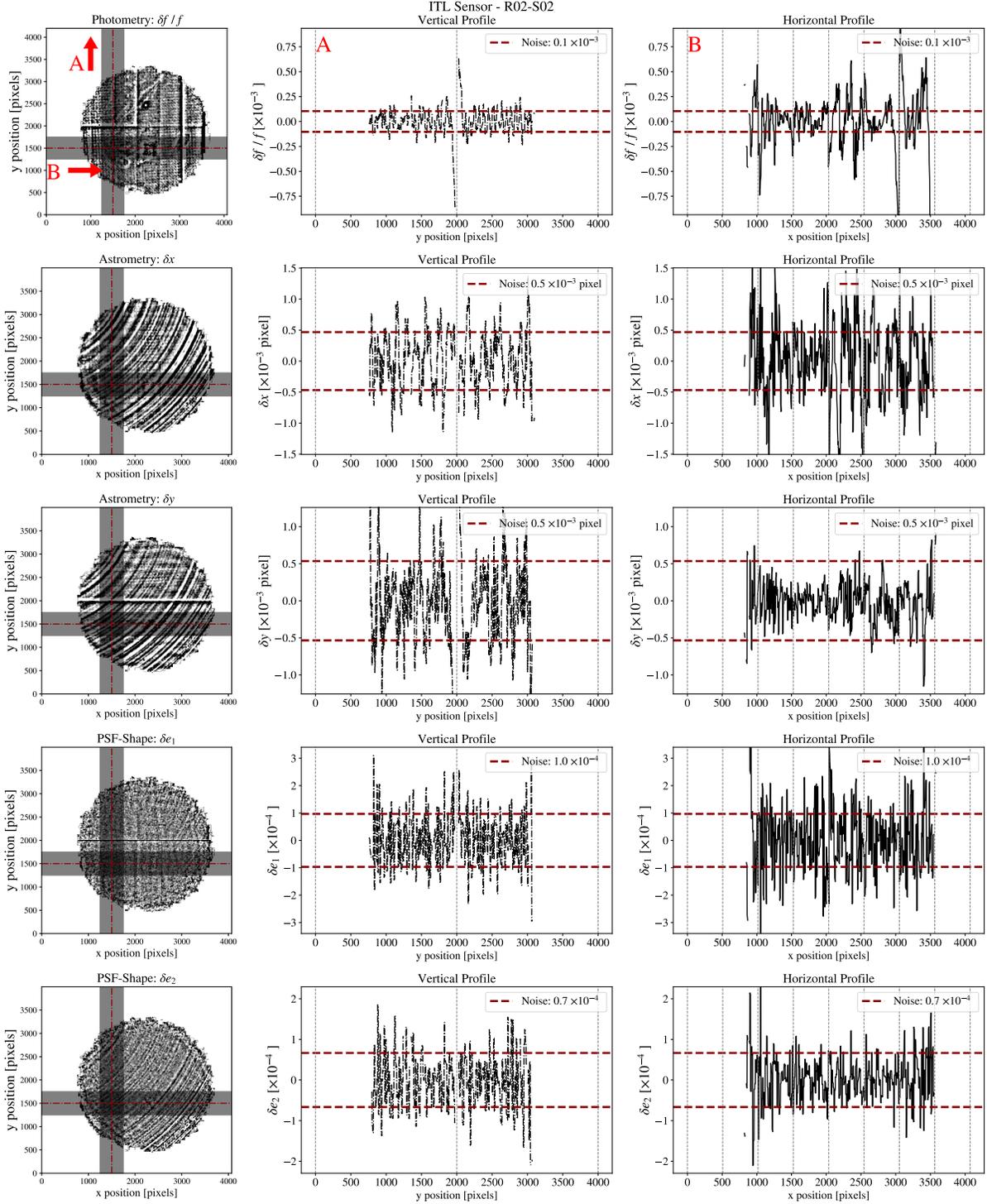}
    \caption{Same as \autoref{fig:signal_profiles} caption. The data was collected during the Run 3 period.}
\end{figure}

\begin{figure*}[!htb]\label{fig:fig3_others}
    \centering
    \includegraphics[page=3, width=1.0\textwidth]{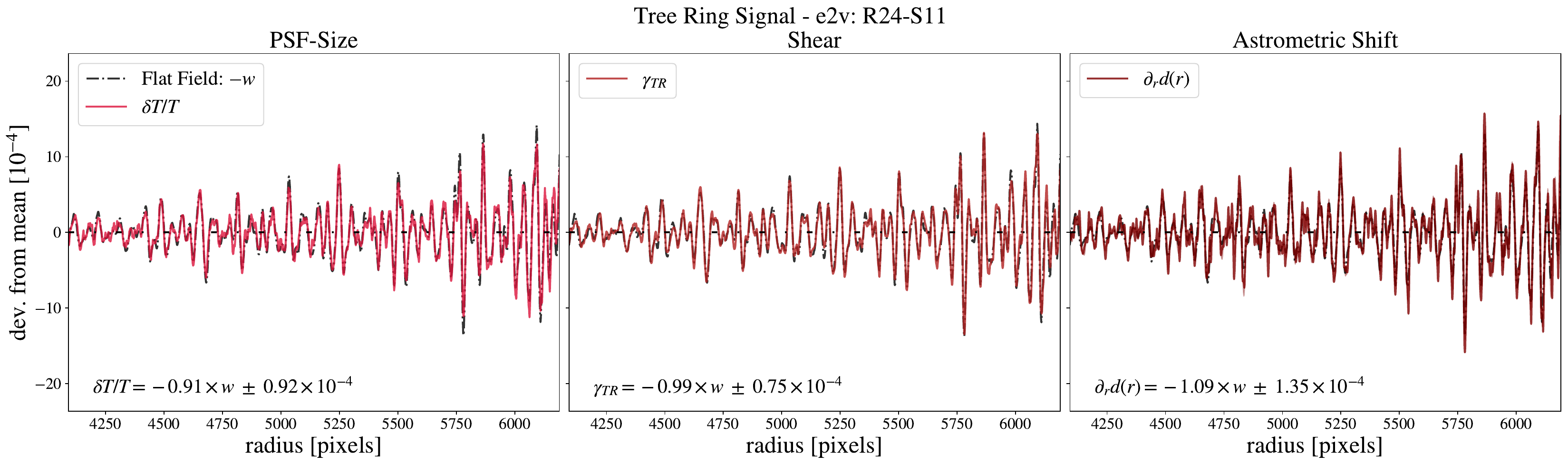}
    \includegraphics[page=2, width=1.0\textwidth]{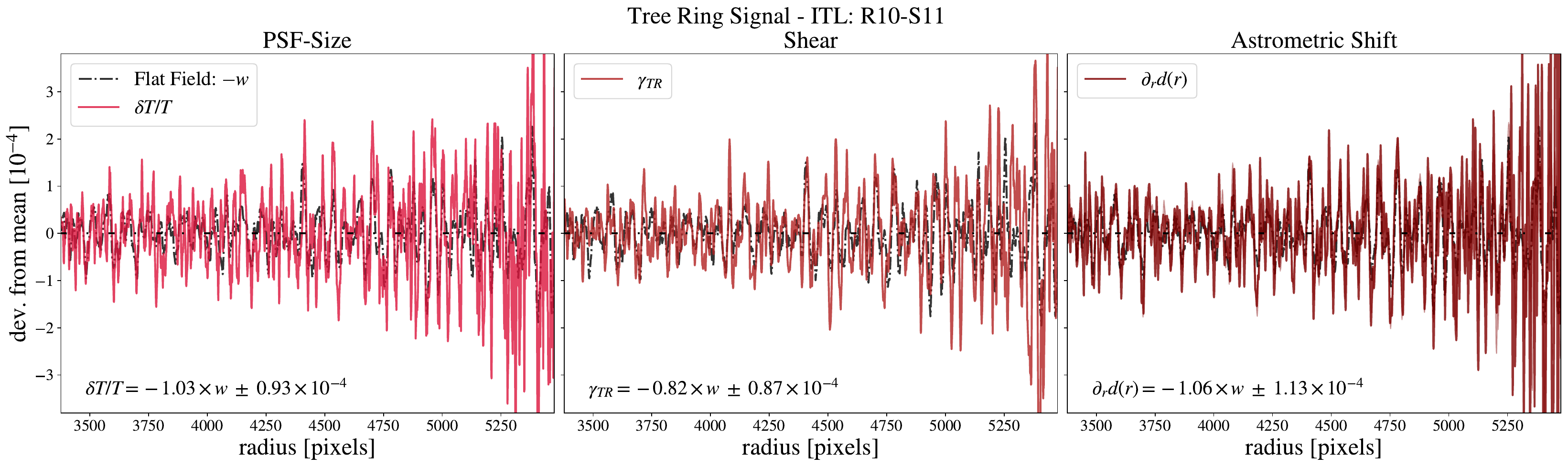}
    \includegraphics[page=5, width=1.0\textwidth]{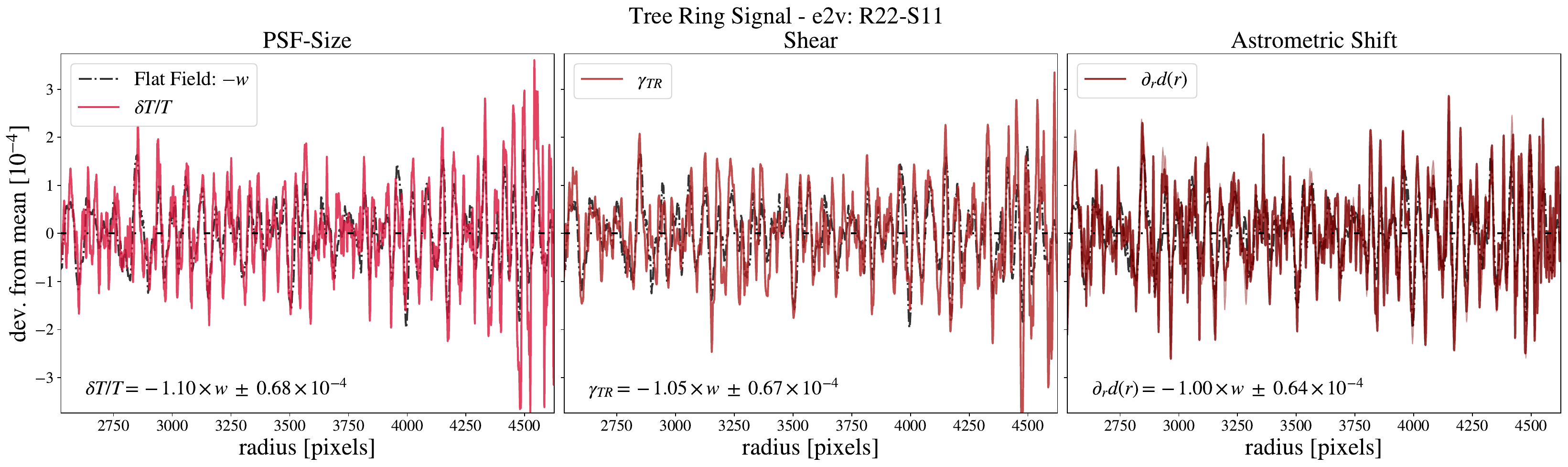}
    \includegraphics[page=6, width=1.0\textwidth]{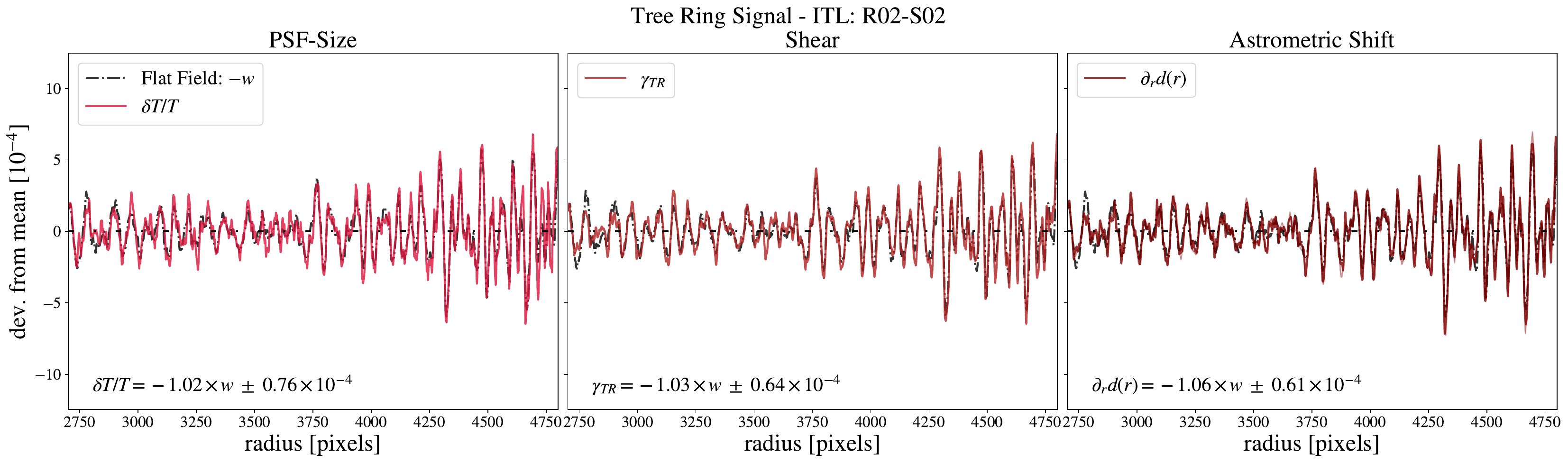}
    \caption{Same as \autoref{fig:tree_ring_signal_panel} caption.}
\end{figure*}

\end{document}